\documentclass[aps,12pt,nofootinbib,superscriptaddress,preprint]{revtex4}
\pdfoutput=1
\usepackage{textcomp}
\usepackage{amssymb}
\usepackage{anysize}
\usepackage{bold-extra}
\usepackage{enumerate}
\usepackage{graphicx}
\usepackage{lscape}
\usepackage{mathrsfs}
\usepackage{multirow}
\usepackage[final]{pdfpages}
\usepackage{rotating}
\usepackage{subfig}
\usepackage{url}
\usepackage{wrapfig}
\usepackage{amsmath}
\usepackage{color}
\usepackage{fancyhdr}
\usepackage{verbatim} 
\def\nn{\nonumber}
\def\bea{\begin{eqnarray}}
\def\eea{\end{eqnarray}}
\def\ba{\begin{eqnarray}}
\def\ea{\end{eqnarray}}
\def\be{\begin{equation}}
\def\ee{\end{equation}}
\def\beq{\begin{equation}}
\def\eeq{\end{equation}}

\def\lsim{\mbox{\raisebox{-.6ex}{~$\stackrel{<}{\sim}$~}}}
\def\gsim{\mbox{\raisebox{-.6ex}{~$\stackrel{>}{\sim}$~}}}

\newcommand{\slashed}{\slash \hspace{-0.23cm}}

\begin{document}
\title{Massive Spin-2 States as the Origin of the Top Quark Forward-Backward Asymmetry}
\preprint{UCSD/PTH 12-02}
\author{Benjam\'{\i}n Grinstein}
\email{bgrinstein@ucsd.edu}
\affiliation{Department of Physics, University of California, San Diego, La Jolla, CA 92093 USA}
\author{Christopher W. Murphy}
\email{cmurphy@physics.ucsd.edu}
\affiliation{Department of Physics, University of California, San Diego, La Jolla, CA 92093 USA}
\author{David Pirtskhalava}
\email{pirtskhalava@physics.ucsd.edu}
\affiliation{Department of Physics, University of California, San Diego, La Jolla, CA 92093 USA}
\author{Patipan Uttayarat}
\email{puttayarat@physics.ucsd.edu}
\affiliation{Department of Physics, University of California, San Diego, La Jolla, CA 92093 USA}
\affiliation{Department of Physics, Srinakharinwirot University, Wattana, Bangkok 10110 Thailand}

\begin{abstract}
We show that the anomalously large top quark forward-backward asymmetry observed by CDF and  D\O\, can naturally be accommodated in models with flavor-violating couplings of a new massive spin-2 state to quarks. Regardless of its origin, the lowest-order couplings of a spin-2 boson to fermions are analogous to the coupling of the graviton to energy/momentum, leading to strong sensitivity of the effects associated with its virtual exchange to the energy scales at hand. Precisely due to this fact, the observed dependence of the asymmetry on the $t\bar t$ invariant mass fits nicely into the proposed framework. In particular, we find a vast parameter space which can lead to the central value for the observed forward-backward asymmetry in the high mass bin.
\end{abstract}

\maketitle
\section{Introduction}
The Standard Model (SM) is a successful theoretical framework for describing elementary particle interactions when confronted with experimental data. 
However, recent observations~\cite{CDFnote:10584, Abazov:2011rq} by the CDF and D\O\, collaborations point to an anomalously large forward-backward asymmetry in $t\bar{t}$ production $\left(A_{FB}^{t\bar{t}}\right)$, significantly exceeding the SM prediction (see Ref.~\cite{Kamenik:2011wt} for a review).  

Perhaps the most intriguing result is CDF's report~\cite{Aaltonen:2011kc} of a rise in $A_{FB}^{t\bar{t}}$ with the invariant mass of the $t\bar{t}$ pair $\left(M_{t\bar{t}}\right)$, 
\ba
A^{t\bar{t}}_\text{{low}} \equiv A_{FB}^{t\bar{t}}(M_{t\bar{t}} \le 450\, \text{GeV}) &= &(-11.6\pm15.3)\%~,  \nn \\ A^{t\bar{t}}_{\text{high}} \equiv A_{FB}^{t\bar{t}}(M_{t\bar{t}} > 450 \,\text{GeV}) &= &(47.5\pm11.4)\% ~.\nn
\ea
In particular, $A^{t\bar{t}}_{\text{high}}$ is almost 3$\sigma$ away from SM prediction (all the relevant measurements, as well as the corresponding SM predictions are collected in Table \ref{tab:sum}).  

This discrepancy invites a new physics (NP) explanation, and many models have been proposed to address the anomalously large $A_{FB}^{t\bar{t}}$.  These models generally involve introducing new scalar ~\cite{Shu:2009xf,Dorsner:2009mq,Dorsner:2011ai,Nelson:2011us,Cheung:2009ch,Patel:2011eh,Blum:2011fa,Stone:2011dn,delaPuente:2011iu} or vector ~\cite{Barger:2010mw,Shelton:2011hq,Grinstein:2011yv,Grinstein:2011dz,Ligeti:2011vt,Tavares:2011zg,Bhattacherjee:2011nr} particles contributing to the $t\bar{t}$ production cross section in the $s$- and/or $t$-channel.  While most of these models can easily raise the theoretical prediction for $A_{FB}^{t\bar{t}}$ to within 1$\sigma$ of the CDF measurement, it has proven to be extremely hard to address the central value of $A^{t\bar{t}}_{\text{high}}$ while being consistent with existing experimental constraints.  

In this work we propose to explore another class of models involving new tensor (spin-2) particles around the weak scale with flavor-violating couplings to quarks. A simple effective field-theoretic (EFT) analysis reveals that the most general lowest-order couplings of a spin-2 state with quark bilinears are rather similar to the general-relativistic couplings of the graviton to energy/momentum. This leads to a strong energy dependence of the effects of virtual exchange of such states, which nicely agrees with the CDF observations. In particular, we show that this framework can accommodate all of the CDF measurements over a wide range of parameter space while being consistent with existing experimental bounds. 

We treat the massive spin-2 particle as a low-energy signature of some unspecified UV physics. Among other possibilities, a low-energy effective theory of this type could arise from a theory of modified gravity~\cite{ArkaniHamed:1998rs, Antoniadis:1998ig,Randall:1999ee, Randall:1999vf,deRham:2010gu, deRham:2010kj} or could describe a spin-2 resonance of a strongly interacting sector not far above the weak scale~\cite{Morningstar:1997ff, Morningstar:1999rf,Chen:2005mg}.

The rest of the paper is organized as follows: in Sec.~\ref{sec:spin2} we describe the relevant couplings of the spin-2 state to quarks.  The existing experimental constraints are examined in Sec.~\ref{sec:constraints}.  We study parameter space for $t\bar{t}$ phenomenology and discuss our results in Sec.~\ref{sec:ttbar}.  Finally, we conclude in Sec.~\ref{sec:con}. 

%
\section{Low-Energy Effective Theory}
\label{sec:spin2}
%
%
%
Studies of low-energy effective field theories for massive spin-2 particles are motivated in various contexts. On one  hand, such theories can be useful for describing spin-2 QCD resonances, such as glueballs, which become long-lived in the limit of large number of colors~\cite{Chen:2005mg}. On the other hand, massive spin-2 states frequently occur in the context of  modified gravity. An incomplete list of examples from the latter category includes models with KK tower of gravitons, such as theories with large \cite{ArkaniHamed:1998rs, Antoniadis:1998ig}  or warped \cite{Randall:1999ee, Randall:1999vf} extra dimensions, as well as the recently discovered class of purely four-dimensional, ghost-free models of massive gravity \cite{deRham:2010gu, deRham:2010kj}.  Ratios of branching ratios to photons and to jets may be used to distinguish between the various possible underlying UV theories~\cite{Fok:2012zk}.

Regardless of the details of the UV theory, any consistent action for a complex, symmetric spin-2 field $h_{\mu\nu}$ with mass $M$ should reduce to the Fierz-Pauli \cite{Fierz:1939ix} form at the linearized level
\begin{equation}
\label{eq:lag}
\mathcal{L}_{FP} = -\frac{1}{2}h^{\dagger}_{\mu\nu}\left(\Box + M^2\right)h^{\mu\nu} + \frac{1}{2}\left.h^{\mu}_{\mu}\right.^{\dagger}\left(\Box + M^2\right)h^{\nu}_{\nu} - h^{\dagger}_{\mu\nu}\partial^{\mu}\partial^{\nu}h^{\rho}_{\rho} + h^{\dagger}_{\mu\nu}\partial^{\mu}\partial^{\rho}h_{\rho}^{\nu} + \text{h.c.}.
\end{equation}
Furthermore, if the field is of gravitational origin, its couplings to matter are usually constrained by the Equivalence Principle to be universal.  However, RS-type models with the SM fields localized differently along the bulk are an important exception to this rule.

In the present work, we will be mostly interested in the implications of a massive, complex, spin-2 boson for the top quark forward-backward asymmetry. Resorting to an EFT approach, we will not make any assumptions about the precise origin of the spin-2 field. Among other possibilities, $h_{\mu\nu}$ could describe a bound state of some strongly coupled sector not far above the electroweak scale, or a non-universally interacting gravitational KK mode in some RS-like theory with complicated localization of matter.  We will not attempt to construct any explicit model along these lines.

At low energies, the sector of the theory describing interactions of $h_{\mu\nu}$ with the quarks consists of operators of various dimension, suppressed by powers of some high scale, denoted by $f$ below.  At the zero derivative order, there is a single coupling of a spin-2 field with a general quark bilinear,
\beq
\mathcal{L}_4 \supset \lambda_{ij} h^{\mu}_{\mu}\bar q_iq_j+\text{h.c.},
\label{0d}
\eeq
where $\lambda$ is a coupling constant, $\{i,j\}$ refer to quark flavor and the possible chirality of $q_i$ has been suppressed for simplicity; the quark fields correspond to the mass eigenstates after electroweak symmetry breaking. This interaction is similar in form to an ordinary Yukawa interaction of a SM singlet scalar; the only difference is in the spin-2 nature of the correlator 
\beq
\langle \left.h^{\mu}_{\mu}\right.^{\dagger}\hspace{-1mm}(k) h^\nu_\nu(-k) \rangle\propto \frac{2-k^2/M^2-(k^2)^2/M^4}{k^2-M^2}.
\eeq 
For the purposes of studying the $t\bar t$ forward-backward asymmetry however, this operator can be expected to lead to effects similar to those of a color-singlet scalar exchange. The consequences of the latter for the $t\bar{t}$ forward-backward asymmetry have been  studied extensively, see for example~\cite{Shu:2009xf}, with the conclusion that it can not generate a large enough asymmetry. Moreover, such inter-generation couplings involving the top quark are constrained not to be too large from the same sign top production cross section at the LHC (see Section~\ref{sec:constraints}); we will thus ignore these operators below\footnote{If $h_{\mu\nu}$ is of gravitational/extra dimensional origin, $f$ represents the quantum gravity scale.  Then the coupling of $h_{\mu\nu}$ to energy-momentum tensor leads to a natural suppression of the Yukawa coupling constant.  In particular, $\lambda\sim m/f$, where $m$ is of the order of the mass of a heavier fermion present in the interaction.}.

At the one derivative order, the most general couplings of $h_{\mu\nu}$ to a fermion bilinear are given by the following expressions, 
\begin{align}
\mathcal{L}_5 &\supset -\frac{1}{f} h_{\mu\nu}\left( S^{\mu\nu} + \eta^{\mu\nu} T\right) + \text{h.c.}, \\
\label{eq:1d}
S_{\mu\nu} &= \frac{i}{4}a^L_{ij}\bar{q}_{Li}\left(\gamma_{\mu}\partial_{\nu} + \gamma_{\nu}\partial_{\mu}\right)q_{Lj} + \frac{i}{4}b^L_{ij}\left(\partial_{\mu}\bar{q}_{Li}\gamma_{\nu}+ \partial_{\nu}\bar{q}_{Li}\gamma_{\mu}\right)q_{Lj} + \left(\text{L} \leftrightarrow \text{R}\right), \\
T &=\bar \lambda^{L}_{i j}\bar q_{Li}\, \slashed{\partial} q_{Lj} + \left(\text{L} \leftrightarrow \text{R}\right),
\label{1d}
\end{align}
with arbitrary complex coefficients $\{a^{L,R},b^{L,R},\bar\lambda^{L,R}\}$.   For external (on-shell) fermions, the interactions in \eqref{1d} can effectively be reduced via the Dirac equation to the non-derivative Yukawa couplings given in \eqref{0d}.  Again, since we are interested in the leading order spin-2 effects on fermion scattering, we will ignore the operators in \eqref{1d} for the remainder of the paper.

\emph{A priori}, there are no constraints on the couplings $a^{L,R}$ and $b^{L,R}$.  However, for a theory in which the spin-2 field is gravitational in nature, $S_{\mu\nu}$ should be related to the energy-momentum tensor. The fact that $h_{\mu\nu}$ interactions are non-universal, as well as couple different generations with each other does not rule out the possible gravitational interpretation of the theory - this can be accommodated e.g. in the framework of RS models with complicated matter localization along the bulk. In such cases the KK gravitons couple to quark \textit{flavor} eigenstates in a diagonal, albeit non-universal way. This, upon rotation to the mass basis, results in the following constraints on the couplings
\beq
a^{L,R}_{ij}=-b^{L,R}_{ij} \equiv g^{L,R}_{ij}.
\eeq
Although we remain completely agnostic about the origin of the spin-2 state, we will take these relations to hold in the analysis to follow; as we show below, restricting the parameter space in this way is already enough for generating the needed amount of asymmetry without running into conflict with other experimental bounds.  Further constraints on the couplings come from the requirement of the invariance of the theory under the SM gauge group.  We will not dwell on making the symmetry manifest, but will keep in mind that it implies some additional relations between the coupling constants.

The key observation to make at this point is that even if the spin-2 state is not associated with any gravitational dynamics, its couplings to fermions are quite similar to the coupling of the graviton to the energy/momentum. This leads to a large sensitivity of the effects associated with the virtual exchange of these states to the energy scales at hand. In particular, we will find that this fact fits with the observed pattern of an increase in $A_{FB}^{t\bar{t}}$  with $t\bar t$ invariant mass.

The range of validity of the low-energy effective theory is an important issue, since it determines the maximum energy to which the analysis performed below can be extrapolated. Given the complete ignorance of the UV completion of low-energy (massive) spin-2 theories, the best we can do is to make an educated guess of the relevant energy scale. For massless general relativity (GR) in 4D, the analysis of graviton loop corrections yields the following expansion parameter \cite{Donoghue:1994dn, Giudice:1998ck},
\beq
\alpha \simeq \left( \frac{E}{4\pi M_{Pl}} \right)^2,
\eeq
where $E$ represents the typical energy scale of a process under consideration, This is in complete analogy to what one finds for low-energy nonlinear sigma models, once $M_{Pl}$ is replaced by the pion decay constant, $f_\pi$. 

The massive spin-2 representation of the Poincar\'e group propagates three additional (one helicity-0 and two helicity-1) degrees of freedom on top of the two helicity-2 modes of the massless theory. In a general nonlinear completion of Fierz-Pauli action, the strong coupling of the helicity-0 mode is usually responsible for the cutoff of the effective theory - in complete analogy to massive non-abelian theories, where the strong coupling of the longitudinal $W$-bosons leads to the necessity of UV completion at a scale $\sim 4\pi m_W/g$ (here $m_W$ and $g$ refer to the $W$-mass and the $SU(N)$ coupling constant). Due to the higher spin structure, non-linear sigma models involving the longitudinal modes of massive spin-2 theories are usually different in nature than those for spin-1\footnote{In particular, a longitudinal scalar mode of a massive spin-2 boson is more strongly coupled than that of a massive spin-1 particle.  Mathematically this can be traced back to the piece in the massive spin-2 propagator which grows fastest with momentum, $$\langle h_{\mu\nu}h_{\alpha\beta}\rangle \supset \frac{k_\mu k_\nu k_\alpha k_\beta}{M^4 (k^2-M^2)} .$$}. This leads to a UV cutoff which is in general sensitive to the nonlinear completion. For example, in theories of massive (four-dimensional) GR with a graviton potential, the cutoff usually comes out to be a certain geometrical mean of the scales $M$ and $f$. A specific class of potentials which avoids the propagation of ghosts in the theory  \cite{deRham:2010gu} leads to a sigma model with higher UV cutoff, compared to theories with a more general potential \cite{ArkaniHamed:2002sp}; some other possible UV/nonlinear completion (e.g. a completion beyong the potential, or one which relaxes the requirement of reproducing four-dimensional GR in the $M\to 0$ limit) can therefore be expected to yield yet more different sigma models. Another possibility is that new physics regulating the low-energy theory in the UV kicks in somewhat below the strong coupling scale\footnote{In extra dimensional examples, one might imagine the higher KK modes softening loop effects, thus providing a completion at intermediate energies - up to the fundamental quantum gravity scale.}, however it does not distort the spin-2 exchange effects up to higher energies.

As already emphasized, below we will not be concerned with the nature of the underlying UV/nonlinear theory and will expect the low-energy description to be valid up to scales somewhat above the scale $\bar f=min\{f/g_i,M\}$ where  $g_i$ collectively denotes all cupling constants in \eqref{0d} - \eqref{1d}.


\section{Existing Experimental Constraints}
\label{sec:constraints}
We now turn our attention to the analysis of the existing experimental constraints on the massive spin-2 model considered in this work. Note that the main motivation of the present work is an illustration of strong energy-dependence of the effects of virtual spin-2 exchange and its possible relevance to observations that exhibit these effects. In this section we are concerned with preliminary estimates of the bounds on the spin-2 parameter space due to various experimental constraints -- just to show that the mechanism can be viable, or even robust. Of course, many additional studies need to be performed (e.g. a closer inspection of LHC bounds or studies of spin correlations of top quark daughters) for a complete phenomenological analysis. Wherever loop contributions are involved, we make the most conservative assumptions for their magnitude just to show that even with these overly restrictive assumptions, there still is a vast parameter space for addressing the $t\bar t$ forward-backward asymmetry.  We leave a more detailed analysis of the experimental bounds for a future study.

Below we determine bounds from LEP, electroweak precision data, same-sign top-quark pair production, $B_d - \overline{B}_d$ mixing and dijet production at the Tevatron.  As can be seen below, the effects of these constraints on the model are mild.  We note however, that a recent analysis~\cite{Berger:2012nw} of the kinematic and dynamical aspects of the relationship between the asymmetries $A^{t\bar{t}}_{FB}$ and $A^{\ell}_{FB}$ measured by D\O\, favors new physics (NP) models that produce more right-handed than left-handed top quarks.  
Even if the left-handed sector is taken to be suppressed, the right-handed sector can still accomodate the asymmetry, as we will show below.  Such a suppression is further motivated by other experimental constraints, such as $B_d-\overline{B}_d$ mixing.  It is also worth noting that the $t\bar{t}$ production cross section, Eqs.~\eqref{eq:T}-\eqref{eq:ST}, of the model at hand is symmetric under the exchange of the left- and  right-handed couplings.

\begin{figure}
\includegraphics[width=0.4\textwidth]{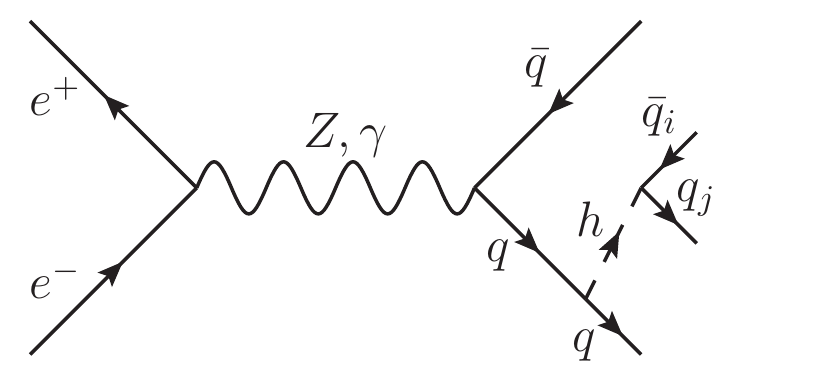}
\caption{One possible diagram for 4-jet production at LEP. The dashed line represents the spin-2 particle exchange.}
\label{fig:lep}
\end{figure}

\subsection{LEP Constraints}
The LEP constraints depend on how the massive spin-2 state couples to electrons and final state quarks.  Since we are only interested in generating a large $A^{t\bar{t}}_{FB}$ at the moment, we can take $g_{ut}\sim\mathcal{O}(1)$ while allowing freedom for the couplings to leptons.  In this scenario therefore, we do not anticipate any bounds from direct production at LEP.  However, due to $SU(2)_W$ symmetry, $g^L_{\{d,b\}} = g^L_{\{u,t\}}$ where $\{u,t\}$ stands for any combination of $u$ and $t$, and a light spin-2 particle could lead to anomalous 4-jet events as shown in Fig.~\ref{fig:lep}.  For examining this bound, we can implement the results in the literature (see e.g. Sec. IV.A. of Ref.~\cite{Grinstein:2011dz} and references therein) in our model. The amplitude for a 4-jet final state in the present model is suppressed by extra factors of $E^2/f^2$ compared to the case of a new scalar or vector field.  Here $E$ denotes a relevant energy scale in the process.
The final operating energy of LEP II is 209 GeV and for the parameter space considered below $f$ is the highest scale in the processes at hand. Even if we conservatively take these suppression factors to be 1 therefore, the bound on the mass of the spin-2 particle from LEP is quite mild, $M \gsim 100$ GeV.

\subsection{Electroweak Precision Tests}
\begin{figure}
\includegraphics[width=0.3\textwidth]{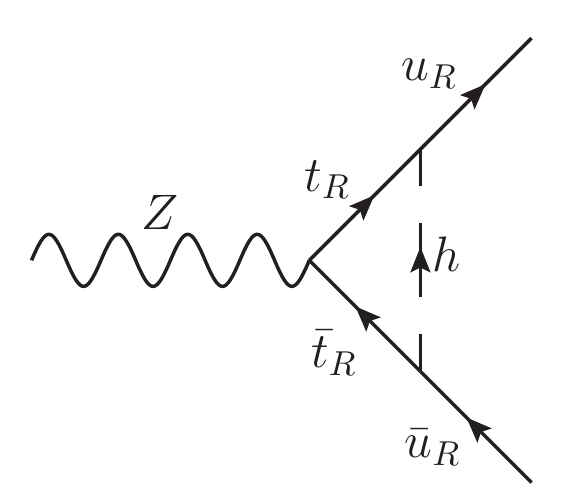}
\caption{The loop contributing to the EW precision observables such as $\Gamma_Z$ and $Q_W$.}
\label{fig:zloop}
\end{figure}

Electroweak precision data (EWPD) can provide strong constraints on models that attempt to explain the $t\bar{t}$ forward-backward asymmetry~\cite{Grinstein:2011gq}.  Corrections to EW precision observables due to the intermediate spin-2 state do not occur at tree level assuming it does not directly couple to the EW gauge bosons.  At the one-loop level there is a contribution to the dimension-4 operator, $C_{Z\bar{u}u} Z_{\mu}\bar{u}\gamma^{\mu}u$, arising from the diagram in Fig. \ref{fig:zloop}.  As shown in Ref.~\cite{Gresham:2012wc}, the most stringent constraint on $C_{Z\bar{u}u}$ comes from atomic parity violation experiments.  The experimental and SM values for $Q_W$ in cesium atoms in the 2010 PDG~\cite{Nakamura:2010zzi} can be turned into a bound on the NP contribution to the coefficient of the dimension-4 operator, $\left|C^{NP}_{Z\bar{u}u}\right| < 1.3\cdot 10^{-3}$.  An estimate of the spin-2 contribution to $Q_W(Cs)$, ignoring left-handed couplings, yields
\begin{equation}
\left|C^{NP}_{Z\bar{u}u}\right| \sim \frac{2e s_w}{3 c_w}\frac{\left|g^R_{ut}\right|^2\left(M^2 + m_t^2\right)}{16\pi^2 f^2} \Rightarrow 
\frac{\left|g^R_{ut}\right|^2\left(M^2 + m_t^2\right)}{ f^2}\lsim 2~.
\end{equation}

\subsection{Single-Top Production}

Single-top, spin-2 production via the reaction $u\,g \rightarrow t\,h_{\mu\nu}$ is PDF enhanced at the LHC and PDF suppressed at the Tevatron relative to spin-2 mediated $t\bar{t}$ production, which has a $q\bar{q}$ initial state.  The phenomenology of spin-2 production can be classified into two categories.  In the first case $h_{\mu\nu}$ is stable on collider time scales, which is predicted in large extra dimensions scenarios.  The decay signature of this reaction - one b-tagged jet, one lepton, and high missing transverse energy (MET) - is not an event that is currently selected in single-top searches at the LHC.  Single-top searches thus far always contain at least 2 jets or 2 leptons.  The other scenario is that the spin-2 particle decays immediately upon production, which is the case in warped extra dimensions scenarios.  In this case, bounds from single-top production can be avoided by making the branching ratio of the spin-2 particle into a $u\bar{u}$ pair small.  As we will shown in Sec.~\ref{sec:dijet} the coupling $g_{uu}$, which controls the size of $Br(h_{\mu\nu} \rightarrow u\bar{u})$, is constrained to be small from dijet bounds.

\subsection{Same-Sign Top-Quark Pair Production}
\begin{figure}
	\subfloat[Tree-level diagram.]{\label{fig:tt_tree}\includegraphics[width=0.38\textwidth]{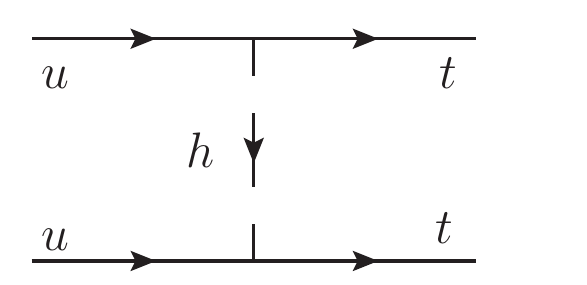}}\qquad
	\subfloat[One of the possible 1-loop diagrams.]{\label{fig:tt_loop}\includegraphics[width=0.45\textwidth]{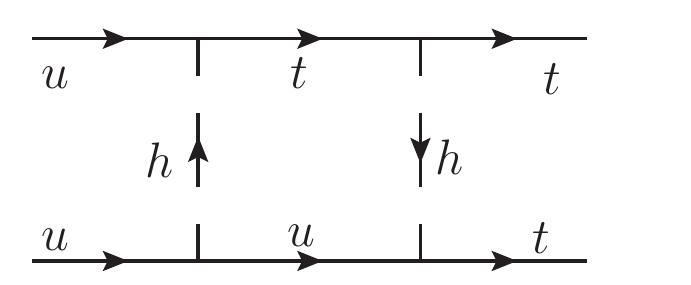}}
	\caption{Feynman diagrams giving rise to the same sign top-quark pair production.}
	\label{fig:tt}
\end{figure}

The $t$-channel models aiming to explain the $A^{t\bar{t}}_{FB}$ asymmetry can be strongly constrained by limits on the same sign top quark pair production at the LHC.  The ATLAS collaboration places limit on the production cross-section at $\sigma_{tt} \le 1.7$ pb~\cite{:2012bb}.  In our model, the production arises from processes shown in Fig.~\ref{fig:tt}.  These diagrams contribute to the coefficient of the effective 4-quark operator responsible for $tt$ pair production, $\mathcal{C}(\bar{t}_R\gamma^\mu u_R)(\bar{t}_R\gamma_\mu u_R)$. The CMS collaboration reports the bound on this coefficient of $\mathcal{C} \le 2.7\text{ TeV}^{-2}$~\cite{Chatrchyan:2011dk}.

\begin{table}[b]
\centering
\begin{tabular}{ |c || c | c | c |  c |  c | c | c |}
\hline 
M [GeV] & 100 & 200 & 300 &400 & 500 & 600 & 700 \\
\hline
$\left|g^R_{tu}\right|$ &$0.05$ & $ 0.04$  & $ 0.03$  & $ 0.03$  & $ 0.03$  & $ 0.03$  & $ 0.02$ \\
\hline
  \end{tabular}
  \caption{95\% CL upper limit on $|g^R_{tu}|$ from a search for same-sign top-quark pair production by the ATLAS collaboration.   $|g^R_{ut}|$ is fixed to be 1, and $f$ is set by requiring that  $A^{t\bar t}_{\text{high}}=0.475$.}
  \label{tab:tt_tree}
\end{table}

The tree-level diagram, Fig.~\ref{fig:tt_tree}, constrains the combination of couplings $g^R_{ut}(g^{R}_{tu})^\ast/f^2$.  We have chosen to taken $g^R_{ut}=1$ in~\eqref{eq:1d} so that the tree-level cross section yields a bound on $g^R_{tu}$, which is given in Table \ref{tab:tt_tree}.

Since the spin-2 propagator contains pieces such as $\frac{k_\mu k_\nu k_\alpha k_\beta}{M^4 (k^2-M^2)}$, loops containing spin-2 particles are highly divergent. The most conservative estimate (i.e. neglecting the possibility of some derivatives acting on the external fermions to reduce the energy-dependence of the finite part of the loop integral) of the one-loop contribution to same-sign top quark production yields
\begin{equation}
\label{eq:sstop}
	\mathcal{C} \sim \frac{\left|g^{R}_{ut}\right|^2 \left|g^{R}_{ii}\right|^2}{16\pi^2}\frac{\hat{s}}{f^4} \frac{\hat{s}^4}{M^8},
\end{equation}
where $i=u$, $t$ and $\hat{s}$ is the partonic center of mass energy.  Since the PDFs drop significantly for the momentum fraction greater than 0.3, we can estimate $\hat{s} = 10\%$ of the LHC running energy, $\hat{s}=700$ GeV. For $f=350$ GeV, $M=500$ GeV and $g^{R}_{ut}=1$, we get $\mathcal{C} \approx 3 \left(g^{R}_{ii}\right)^2 \text{ TeV}^{-2}$.  This leads to the bound $g^{R}_{ii}\lesssim 0.9$.  Note that the bounds get weaker for a larger value of $f$ and/or $M$. 

The bound from the non-derivative interactions in Eq.~\eqref{0d} can be easily obtained in a similar fashion.  But since this coupling doesn't play much role in our analysis of $A_{FB}^{t\bar{t}}$, we will not dwell on the value of this bound.  Put differently, we can take the coupling $\lambda_{ij}$ to be negligible while still producing a large $A_{FB}^{t\bar{t}}$.


Single-top, spin-2 production where the spin-2 particle immediately decays into $t\bar{u}$ or a $t\bar{t}$ pair can also contribute to same-sign top production.  We aproximate $\sigma(u\,g \rightarrow t\,t\, j) \approx \sigma (u\,g \rightarrow t\,h_{\mu\nu})\times \left( Br(h_{\mu\nu} \rightarrow t\,\bar{t}) +  Br(h_{\mu\nu} \rightarrow t\,\bar{u}) \right)$ as we are only interested in producing a quick estimate of the cross section.  The ATLAS collaboration~\cite{:2012bb} imposes a cut, $|\eta| < 2.5$, when selecting lepton and jet candidates assoicated with same-sign top production.  We imposed this cut when calculating the spin-2 contribution to this cross section.  We neglect $Br(h_{\mu\nu} \rightarrow t\,\bar{u})$ since this branching ratio is proportional to $|g_{tu}|^2$, which is constrained to be small from same-sign top production in other channels.  For $M = 400$ GeV and $f = 1$ TeV, the ATLAS bound on same-sign top production, $\sigma_{tt} < 1.7$ pb, yields the constraint, $|g_{ut}|^2 Br(h_{\mu\nu} \rightarrow t\,\bar{t}) \lsim 0.97$.  

\subsection{$B_d-\overline{B}_d$ Mixing}
\begin{figure}
\includegraphics[width=0.35\textwidth]{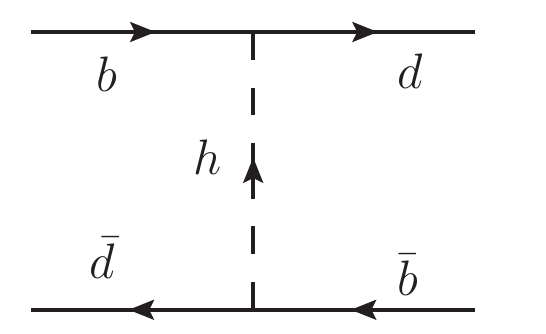}
   \caption{A tree-level contribution to $B_d-\overline{B_d}$ mixing.}
\label{fig:bbbar}
\end{figure}

Non-zero off-diagonal couplings can lead to flavor changing neutral currents (FCNCs).
Here we focus on $B_d-\overline{B}_d$ mixing as the most restrictive bounds on $g^{L}_{ut}$ are expected to come from this process.  The spin-2 contributions to $B_d$ mixing can be described by the following four-quark operator
\begin{align}
\label{eq:ops}
	Q_{1} &= \Big(\bar{d}_{L}\gamma_{\mu}b_{L}\Big) \Big(\bar{d}_{L}\gamma^{\mu}b_{L} \Big).
\end{align}
In general, other operators contribute to $B_d$ mixing as well.  The coefficient of the operator in Eq.~\eqref{eq:ops} is constrained to be smaller than $\mathcal{O}(10^{-11})\text{ GeV}^{-2}$~\cite{Bona:2007vi}. These bounds constrain the couplings relevant for $A^{t\bar{t}}_{FB}$ in our model, in particular there is a constraint on $g^L_{ut}$.
This is due to the fact that by $SU(2)_L$ symmetry, $g^L_{db} = g^L_{ut}$.  
The contribution to this operator arises at tree-level from Fig.~\ref{fig:bbbar}.
A quick dimensional analysis estimate reveals that the contribution from such a diagram is
\begin{equation*}
	\frac{g^L_{ut} \left(g^{L}_{tu}\right)^\ast} {f^2} \frac{m_b^2}{M^2} \approx 4g^L_{ut} \left(g^L_{tu}\right)^\ast 10^{-10}\, \text{GeV}^{-2},
\end{equation*}
where we estimate $m_b/M\sim10^{-2}$ and take $f=500$ GeV.  Thus we see that for $g^L_{ut}(g^L_{tu})^\ast \lesssim \mathcal{O}(0.01)$, the constraints from $B_d-\overline{B}_d$ mixing can easily be satisfied.  As in the case of same sign top-quark pair-production, the constraint gets weaker for larger values of $f$. Similar constraints hold for the right-handed couplings.  However, there is no symmetry relating $g^R_{db}$ to $g^R_{ut}$.  Thus there is more freedom available in the right-handed sector to address the top forward-backward asymmetry.

\subsection{Tevatron Dijet and Top Width Constraints}
\label{sec:dijet}
\begin{table}[b]
\centering
\begin{tabular}{ |c || c | c | c |  c |  c | c | c |}
\hline 
 $M$ [GeV] &300 & 400 & 500 &700 &900 &1100 &1300   \\ \hline
 $\left|g^R_{uu}\right|$& 0.08 & 0.15& 0.16& 0.18& 0.17& 0.12& 0.09 \\
\hline
  \end{tabular}
  \caption{95\% CL upper limit on $|g^R_{uu}|$ from dijet constraints at the Tevatron.  $|g^R_{ut}|$ is fixed to be 1, and $f$ is set by requiring that  $A^{t\bar t}_{\text{high}}=0.475$.}
  \label{tab:dijet}
\end{table}
As emphasized above, the derivative couplings of the spin-2 field to light quark pairs grow with energy, which can lead to strong constraints on the couplings at large invariant mass, $M_{jj}$.  Following Ref.~\cite{Grinstein:2011dz}, we obtain the Tevatron dijet bounds from the CDF 95\% CL upper limits on the product of an RS graviton ($G^{\star}$) production cross section $\times$ its branching ratio to dijets ($\mathcal{B}$) $\times$ acceptance ($\mathcal{A}$); see Table 1 of Ref.~\cite{Aaltonen:2008dn}. The results with $\mathcal{B}\cdot\mathcal{A} = 1$ are collected in Table~\ref{tab:dijet}.    The CDF analysis uses the RS parameter $k / \bar{M}_{Pl} = 0.1$, which translates into $ M \approx 0.383 {f}$ in the notation of this paper (with the convention that the largest dimensionless coupling is set to one).  The dijet cross section in the present model is therefore related to the RS graviton cross section by $\sigma^{NP}_{jj} = C^4\,\sigma_{G^{\star}}$, where $ C = \left|g^R_{uu}\right| M / 0.383 f$.  To obtain a meaningful bound, $f$ is taken to be of the value required to produce an asymmetry in the high mass bin of 47.5\% with $\left|g^R_{ut}\right| = 1$.  $\sigma^{NP}_{jj}$ is estimated to be the $s$-channel NP cross section.  $\sigma^{NP}_s$, see Eq.~\eqref{eq:S}, includes additional terms due to a finite top-quark mass; $F_i(x,\,0)$ should be used be for light quark production cross sections as opposed to $F_i(x,\,y)$.  This is an overestimate of the actual dijet cross-section, which is produces a conservative estimate of the bound.

$g^R_{uu}$ does not play a role in generating a large asymmetry.  However, $g^R_{uu}$ along with $g^R_{tt}$ are important for increasing the total cross section from the SM value up to what is measured at the Tevatron (see Table~\ref{tab:sum} for the relevant measurements and the corresponding SM predictions).  A smaller value of $g^R_{uu}$ requires a larger value of $g^R_{tt}$ to produce the same cross section.  A larger $g^R_{tt}$ and a smaller $g^R_{uu}$ could be expected in an RS model; localization of the top quark close to the IR brane leads to a large coupling, while the light quark couplings are relatively suppressed as they are localized in the bulk closer to the UV brane.

The dijet invariant mass and angular distributions measured the LHC may very well be more constraining than the Tevatron measurements.  However, we do not consider LHC dijet data because these measurements only constrain the coupling $g_{uu}$ in this model.  As previously noted, $g_{uu}$ plays no role in generating an asymmetry so constraining this coupling to be smaller does not affect the goal of this work.  LHC dijet measurements will be important for constraining a more complete model, and we leave this work for a future publication.

When $M < m_t$, the non-standard top quark decay $\Gamma\left(t \rightarrow u\,h_{\mu\nu}\right)$ is allowed,
\begin{equation}
\Gamma_t^{NP} = \frac{\left|g_{ut} + g_{tu}\right|^2 m_t^7}{12 \pi f^2 M^4}\left(1 - \frac{M^2}{m_t^2}\right)^4\left(2 + 3\frac{M^2}{m_t^2}\right).  
\end{equation}
The recent D\O\, measurement~\cite{Abazov:2012vd}, $\Gamma_t = \left(2.00^{+0.47}_{-0.43}\right)$ GeV, constrains how large $\left|g_{ut}\right| / f$ can be in the low mass region; see Figs.~\ref{fig:chi1} and~\ref{fig:chiC} for bounds on the spin-2 parameter space from $\Gamma_t$.

\section{Top Quark Forward-Backward Asymmetry}
\label{sec:ttbar}

\subsection{Calculation of the Cross Section}

\begin{figure}
\includegraphics[width=0.65\textwidth]{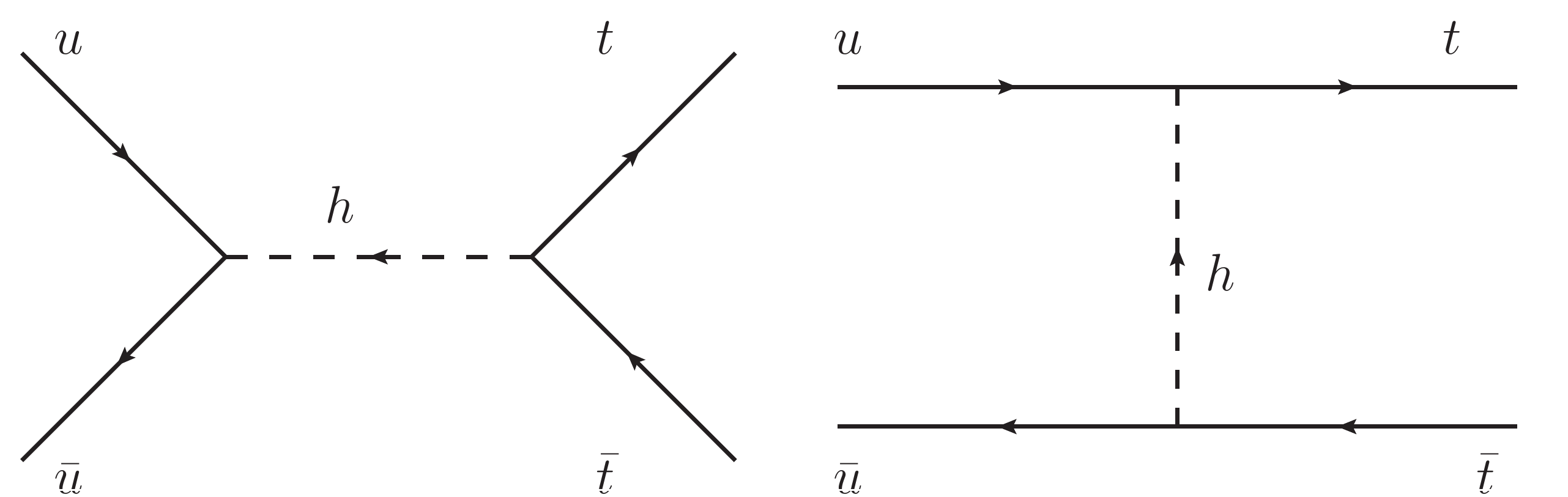}
\caption{Most important spin-2 contributions to $t\bar t$ production at the Tevatron.}
\label{fig:hexchange}
\end{figure}

In the present section we study the effects of an intermediate massive spin-2 state on the top quark forward-backward asymmetry at the Tevatron.  Calculations of the virtual exchange of a massive spin-2 graviton in the context of theories with large extra dimensions were done by Giudice \textit{et al}. in~\cite{ Giudice:1998ck}.  However, the results of Ref.~\cite{ Giudice:1998ck} should be applied to collider phenomenology with care as they are only valid in the limit (in our notation) $f^2 \sim M^2 \gg \hat{s} \gg m_t^2$.  Here we extend the results of~\cite{ Giudice:1998ck} by assuming there is no hierarchy between the aforementioned  scales.

The weighted average (the average over initial spins and colors and the sum over final spins and colors) of the amplitude squared for the $q\bar{q} \rightarrow t\bar{t}$ scattering from a spin-2 $t$-channel exchange is
\begin{align}
\label{eq:T}
\langle\left|\mathcal{M}_t\right|^2\rangle &= \frac{\hat{s}^4}{128f^4\left(\left(\hat{t} - M^2\right)^2 + \Gamma^2 M^2\right)} \left[C_1^2\left(F_1 + \frac{m_t^2}{M^2}F_2+ \frac{m_t^4}{M^4}F_3\right)\right. \nn \\
&\left. + \frac{C_2^2}{18}\left(F_4 + \frac{m_t^2}{M^2}F_5 + \frac{m_t^4}{M^4}F_6 + \frac{m_t^6}{M^6}F_7 + \frac{m_t^8}{M^8}F_8\right)\right],
\end{align}
where $\hat{s},\,\hat{t}$ are the Mandelstam variables in the parton center-of-momentum frame, $q = \{u,\,c\}$ is assumed to be massless, and $\Gamma$ denotes the width of the spin-2 resonance.  The functions $F_i$ are polynomials in $x \equiv \hat{t}/\hat{s}$ and $y \equiv m_t^2/\hat{s}$, and are defined in Appendix~\ref{sec:ap}. The $C_i$'s are combinations of couplings and are also given in Appendix~\ref{sec:ap}.  

The interference of the $t$-channel spin-2 exchange with the SM leading order (LO) gluon exchange gives
\begin{equation}
\label{eq:Tint}
\langle2\,\text{Re}\left(\mathcal{M}_t\mathcal{M}_{SM}^{\ast}\right)\rangle = \frac{2\pi\alpha_s \hat{s}^2\left(\hat{t} - M^2 \right)}{27f^2\left(\left(\hat{t} - M^2\right)^2 + \Gamma^2 M^2\right)} \,C_2 \left(F_9 + \frac{m_t^2}{M^2}F_{10}+ \frac{m_t^4}{M^4}F_{11}\right),
\end{equation}
while the weighted average of the amplitude squared for the $q\bar{q} \rightarrow t\bar{t}$ scattering from a spin-2 $s$-channel exchange is
\begin{equation}
\label{eq:S}
\langle\left|\mathcal{M}_s\right|^2\rangle = \frac{\hat{s}^4}{128f^4\left(\left(\hat{s}- M^2\right)^2 + \Gamma^2 M^2\right)} \left(C_3\,F_{12} + C_4\,F_{13} + C_5\,F_{14} \right),
\end{equation}
where $q=\{u,\,d,\,s,\,c,\,b\}$.  The exchange of any color-singlet particle in the $s$-channel can not interfere with color-octet gluon exchange in the SM.  However, $s$-channel spin-2 exchange can interfere with the exchange of a spin-2 particle in the $t$-channel with $q=\{u,\,c\}$,
\begin{align}
\label{eq:ST}
\langle 2\,\text{Re}\left(\mathcal{M}_t\mathcal{M}_{s}^{\ast}\right)\rangle &= \frac{\hat{s}^4\left(\left(\hat{s} - M^2\right)\left(\hat{t} - M^2\right) + \Gamma^2 M^2\right)}{1152f^4\left(\left(\hat{s} - M^2\right)^2 + \Gamma^2 M^2\right)\left(\left(\hat{t} - M^2\right)^2 + \Gamma^2 M^2\right)} \\
&\times\left[C_6 \left(F_{15} + \frac{m_t^2}{M^2}F_{16}+ \frac{m_t^4}{M^4}F_{17}\right) + C_7 \left(F_{18} + \frac{m_t^2}{M^2}F_{19}+ \frac{m_t^4}{M^4}F_{20}\right)\right]. \nn
\end{align}
Our results are consistent with what was found in Ref. \cite{ Giudice:1998ck}\footnote{For example, $F_{12}(x,\,0) = G_4(x)$ and $F_4(x,0)/18 + F_{15}(x,0)/6 = G_{11}(x)$, where $G_i(x)$ are given in the appendix of~\cite{ Giudice:1998ck}.}.

\subsection{Tevatron Measurements and SM Predictions}
\label{sec:teva}
We start out by reviewing the recent observations of the anomalously large top quark forward-backward asymmetry, $A^{t\bar{t}}_{FB}$.  The experimental evidence for contributions to $A^{t\bar{t}}_{FB}$ from physics beyond the SM is as follows.  The CDF collaboration measured~\cite{CDFnote:10584} the asymmetry to be 
$(20.0 \pm 7.0)\%$. 
A recent D\O\, analysis~\cite{Abazov:2011rq} yielded the value
$A^{t\bar{t}}_{FB} = 19.6^{+6.2}_{-6.5}\%$,
in good agreement with the CDF measurement.  D\O\, also reports a forward-backward asymmetry based on the rapidity of the leptons from top quark decays of $A^{\ell}_{FB} = (15.2 \pm 4.0)\%$ compared with the small SM value $(2.1 \pm 0.1)\%$ calculated using \texttt{MC@NLO}.  All uncertainties have been added in quadrature.  In addition, the CDF collaboration reports~\cite{Aaltonen:2011kc} that the asymmetry rises with the invariant mass of the $t\bar{t}$ system, with $A^{t\bar{t}}_{\text{high}} \equiv A^{t\bar{t}}_{FB}\left(M_{t\bar{t}} > 450\,\text{GeV}\right) = (47.5 \pm 11.4)\%$, and $A^{t\bar{t}}_{\text{low}} \equiv A^{t\bar{t}}_{FB}\left(M_{t\bar{t}} \le 450\,\text{GeV}\right) = -(11.6 \pm 15.3)\%$.  

\begin{table}[b]
\centering
 \begin{tabular}{  c | c  c }
  \hline \hline
 Observable & Measurement & SM prediction~\cite{Manohar:2012rs} \\ \hline
$A^{t\bar{t}}_{FB}$ & $(20.0 \pm 4.7)\%$~\cite{Grinstein:2011dz} & $9.3^{+2.7}_{-2.5}\%$ \\
$A^{t\bar{t}}_{\text{high}}$ & $(47.5 \pm 11.4)\%$~\cite{Aaltonen:2011kc} & $14.1^{+3.2}_{-2.6}\%$ \\
$A^{t\bar{t}}_{\text{low}}$ & $-(11.6 \pm 15.3)\%$~\cite{Aaltonen:2011kc} & $5.4^{+0.9}_{-0.6}\%$ \\
$\sigma_{t\bar{t}}$ & $(8.5 \pm 0.9) \text{ pb}~\text{\cite{Aaltonen:2010hza}}$ & $6.59^{+0.24}_{-0.40}$ pb \\
   \hline \hline
  \end{tabular}
  \caption{Measurements and predictions for observables in $t\bar{t}$ production at the Tevatron.}
  \label{tab:sum}
\end{table}

Despite recent improvements in the SM calculations, the asymmetry in the high mass bin is still close to three standard deviations away from the SM value.  The central value of a next-to-leading order plus next-to-next-to-leading logarithm (NLO+NNLL) QCD calculation of $A^{t\bar{t}}_{FB}$ is $7.3^{+1.1}_{-0.7}\, \%$~\cite{Ahrens:2011mw}.  Recently calculated electroweak Sudakov (EWS) corrections enhance the QCD asymmetry by a factor of 1.041 to $7.7\%$~\cite{Manohar:2012rs}, while fixed order electroweak contributions add an additional $1.6\%$ to the asymmetry~\cite{Hollik:2011ps}.  
The overlap between the EWS corrections and the fixed order EW contributions is estimated in Ref.~\cite{Manohar:2012rs} to be $\sim 0.5\%$.  This yields a total SM prediction of $A^{SM}_{FB} = 9.2^{+2.8}_{-2.6}\,\%$.  Similarly, combining the QCD predictions of Ref.~\cite{Ahrens:2011uf} (which use MSTW2008 PDFs~\cite{Martin:2009iq}) and EW effects calculated in Refs.~\cite{Manohar:2012rs, Hollik:2011ps}, the total SM prediction for $A^{SM}_{FB}$ in the low and high mass bins is $5.4^{+0.9}_{-0.6}\,\%$ and $14.1^{+3.2}_{-2.6}\,\%$ respectively.  The $2.8\sigma$ deviation from the SM in the high mass bin may be taken as a signal of new physics (NP).

The total cross section for $t\bar{t}$ production was recently measured~\cite{Aaltonen:2010hza} by CDF to be $\sigma_{t\bar{t}}= (8.5\pm 0.9)$ pb. This measurement is consistent with the value reported~\cite{Abazov:2011mi} by D$\O$, $\sigma_{t\bar{t}}= 7.78^{+0.77}_{-0.64}$ pb.  Cacciari \textit{et al}.~\cite{Cacciari:2011hy} calculated the total cross section at approximate NNLO QCD to be $6.722^{+0.243}_{-0.410}$ pb.  The EWS correction factor for this observable is $\mathcal{R}_t = 0.98$~\cite{Manohar:2012rs}.  For $M_{t\bar{t}} \le 450$ GeV the NLO+NNLL SM prediction for $\sigma^{SM}$ is 4.23 pb, while for $M_{t\bar{t}} > 450$ GeV $\sigma^{SM} = 2.40$ pb~\cite{Grinstein:2011dz}.  $\mathcal{R}_t$ in these bins is 0.985 and 0.973 respectively\footnote{We thank Mike Trott for the computation of the EWS correction factors for $\sigma_{t\bar{t}}$ in the low and high mass bins.}.  The measured and predicted values for $t\bar{t}$ observables at the Tevatron are summarized in Table~\ref{tab:sum}.


\subsection{New Physics Results}
Here we show that tree-level exchanges of a massive spin-2 particle can contribute significantly to $A^{t\bar{t}}_{FB}$, ameliorating the tension between measurements at the Tevatron and SM predictions. Following Ref.~\cite{Grinstein:2011dz}, we define a partonic level asymmetry,
\begin{equation}
A^{NP+SM}_{FB} = \frac{\sigma^{NP}_F - \sigma^{NP}_B}{\left(\sigma^{SM}\right)_{LO} + \sigma^{NP}} + A^{SM}_{FB} \frac{\sigma^{SM}}{\sigma^{SM} + \sigma^{NP}},
\end{equation}
which is to be compared against the binned partonic asymmetries reported in~\cite{Aaltonen:2011kc}.  For later convenience, we define
\begin{equation}
A^{NP}_{FB} = \frac{\sigma^{NP}_F - \sigma^{NP}_B}{\left(\sigma^{SM}\right)_{LO} + \sigma^{NP}}
\end{equation}
as the NP contribution to the top-quark forward-backward asymmetry.
 We use state-of-the-art predictions for the SM quantities $A^{SM}_{FB}$ and $\sigma^{SM}$, and LO predictions for the NP corrections.  The partonic NP cross sections are convoluted into hadronic cross sections using NLO MSTW2008 parton distribution functions (PDFs)~\cite{Martin:2009iq}.  The factorization and renormalization scales are taken to be $\mu = m_t = 173.1$ GeV, while $\alpha_s$ is set by the MSTW fit value: $\alpha_s\left(M_Z\right) = 0.12018$.  The width of the spin-2 state is taken to be a tenth of its mass, $\Gamma = M / 10$.  We take all of the couplings to be real, and ignore contributions from $F_{14}(x,y)$ as it is numerically negligible.

\begin{figure}
  \centering
  \includegraphics[width=0.7\textwidth]{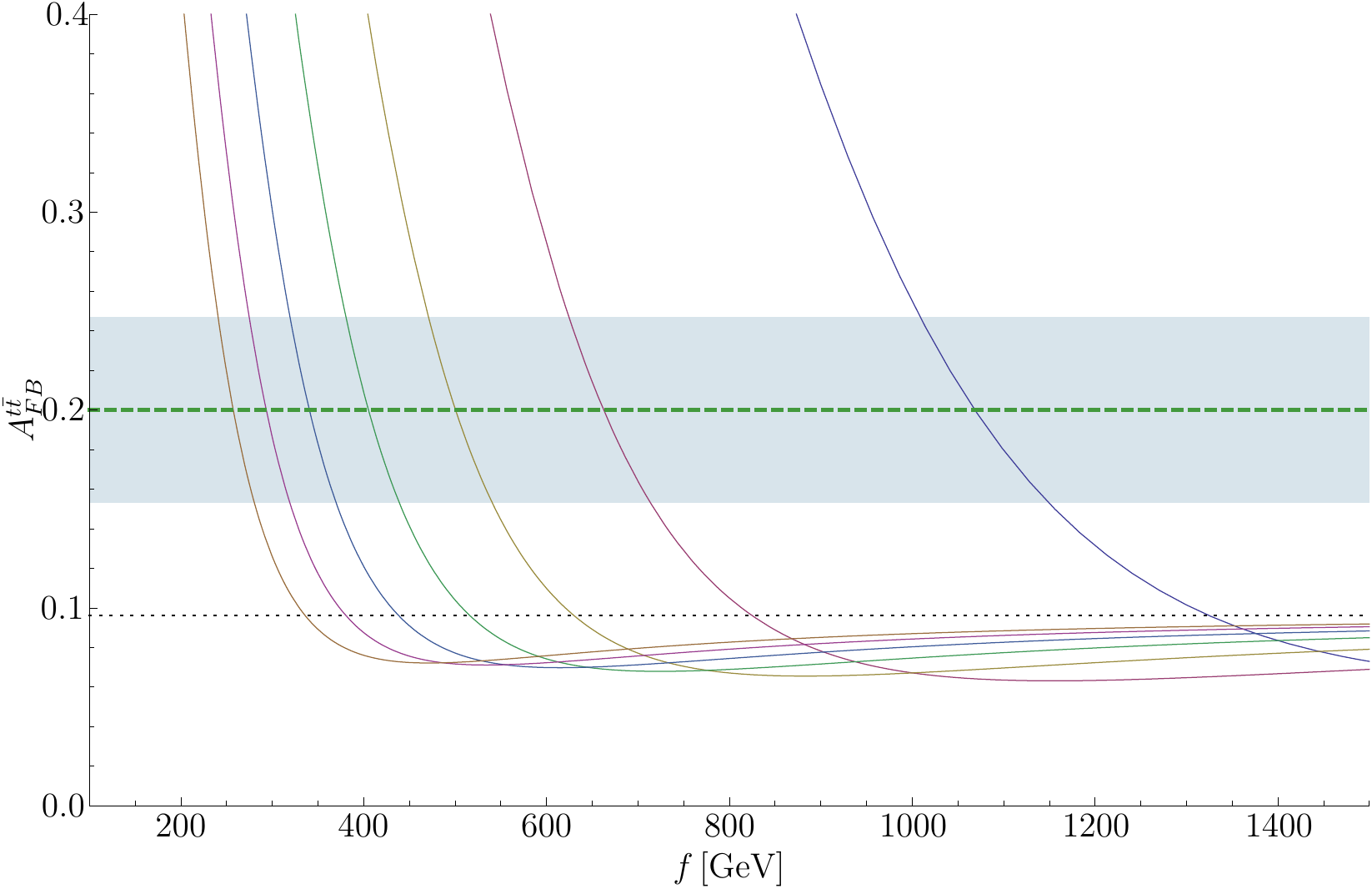}
    \caption{$A^{t\bar{t}}_{FB}$ vs. $f$.  $\left|g^R_{ut}\right| = 1$ and all other couplings are zero.  The thick, dashed line is the central value as measured at the Tevatron with $1\sigma$ error bands.  The dotted line is the SM prediction.  From left to right, $M$ decreases from 700 GeV to 100 GeV in steps of 100 GeV.}
    \captionsetup{justification=raggedright}
 \label{fig:1}
   \end{figure}

The total asymmetry as a function of $f$ is shown in Fig.\ref{fig:1}. Here $\left|g^R_{ut}\right| = 1$ and we have set all other couplings to zero.  The thick, dashed line is the central value as measured at the Tevatron with $1\sigma$ error bands.  The dotted line is the SM prediction.  From left to right (at the top of the plot), $M$ decreases from 700 GeV to 100 GeV in steps of 100 GeV.  For intermediate values of $f$, the destructive interference with the SM exceeds the pure NP contribution, decreasing the asymmetry.   As expected, the NP decouples for large $f$.

\begin{figure}
  \centering
    \subfloat[$t$-channel NP]{\label{fig:T}\includegraphics[width=0.5\textwidth]{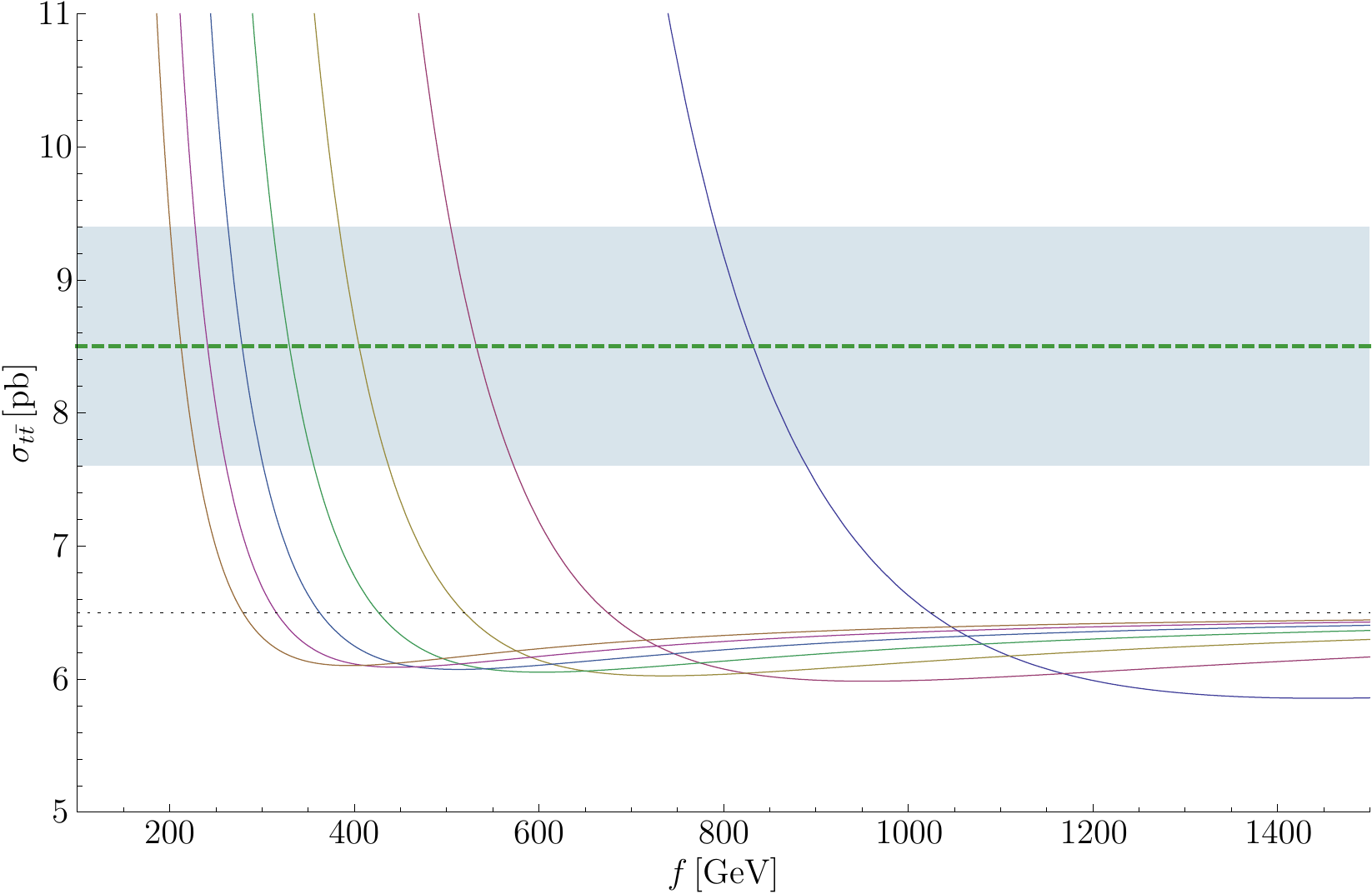}}
    \subfloat[$s$-channel NP]{\label{fig:S}\includegraphics[width=0.5\textwidth]{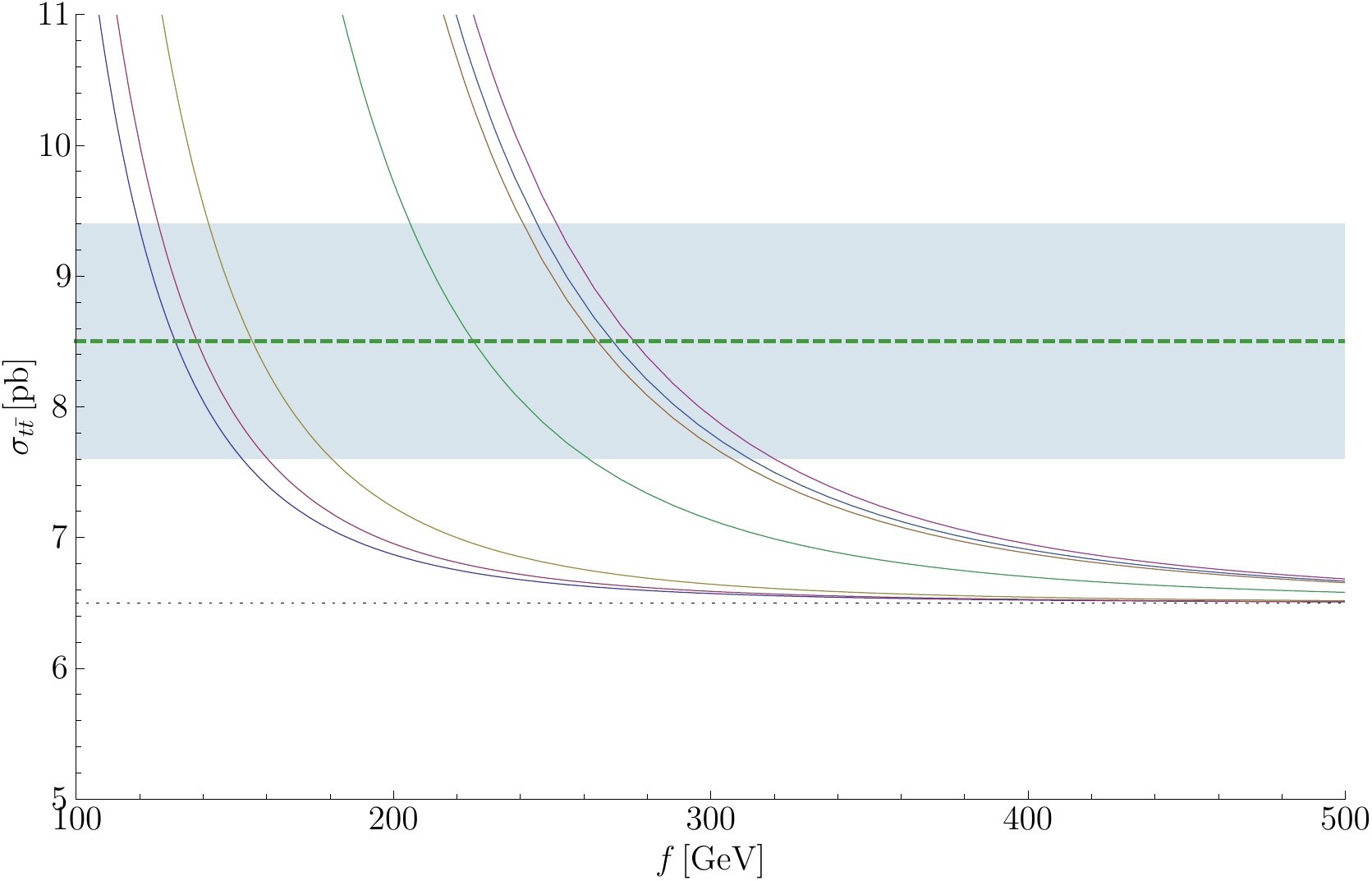}}
    \caption{$\sigma_{t\bar{t}}$ vs. $f$.  $t$-channel NP is shown in Fig.~\ref{fig:T} with $\left|g^R_{ut}\right| = 1$ and all other couplings set to zero.  Fig.~\ref{fig:S} shows $s$-channel NP with the only non-zero coupings being $\left|g^R_{uu}\right| = \left|g^R_{tt}\right| = 1$.  In Fig.~\ref{fig:T}, $M$ decreases from 700 GeV to 100 GeV from left to right in steps of 100 GeV.  In Figure~\ref{fig:S} however, the ordering is not as simple, $M = \{100,\,200,\,300,\,400,\,700,\,500,\,600\}$ from left to right.}
    \captionsetup{justification=raggedright}
 \label{fig:2}
   \end{figure}

Fig.~\ref{fig:2} shows the effects of NP on $\sigma_{t\bar{t}}$ as a function of $f$.  The cross section for $t$-channel NP is shown in Fig.~\ref{fig:T} with $\left|g^R_{ut}\right| = 1$ and all the other couplings set to zero. While Fig.~\ref{fig:S} shows the effect of $s$-channel NP as a function of $f$ with the only non-zero coupings being $\left|g^R_{uu}\right| = \left|g^R_{tt}\right| = 1$.  In Fig.~\ref{fig:T}, $M$ again monotonically decreases from 700 GeV to 100 GeV from left to right in steps of 100 GeV.  In Figure~\ref{fig:S} however, the ordering is not as simple, $M = \{100,\,200,\,300,\,400,\,700,\,500,\,600\}$  from left to right.

\begin{figure}
  \centering
   \subfloat[$g^R_{uu} = g^R_{tt} = 0$]{\label{fig:chi1}\includegraphics[width=0.5\textwidth]{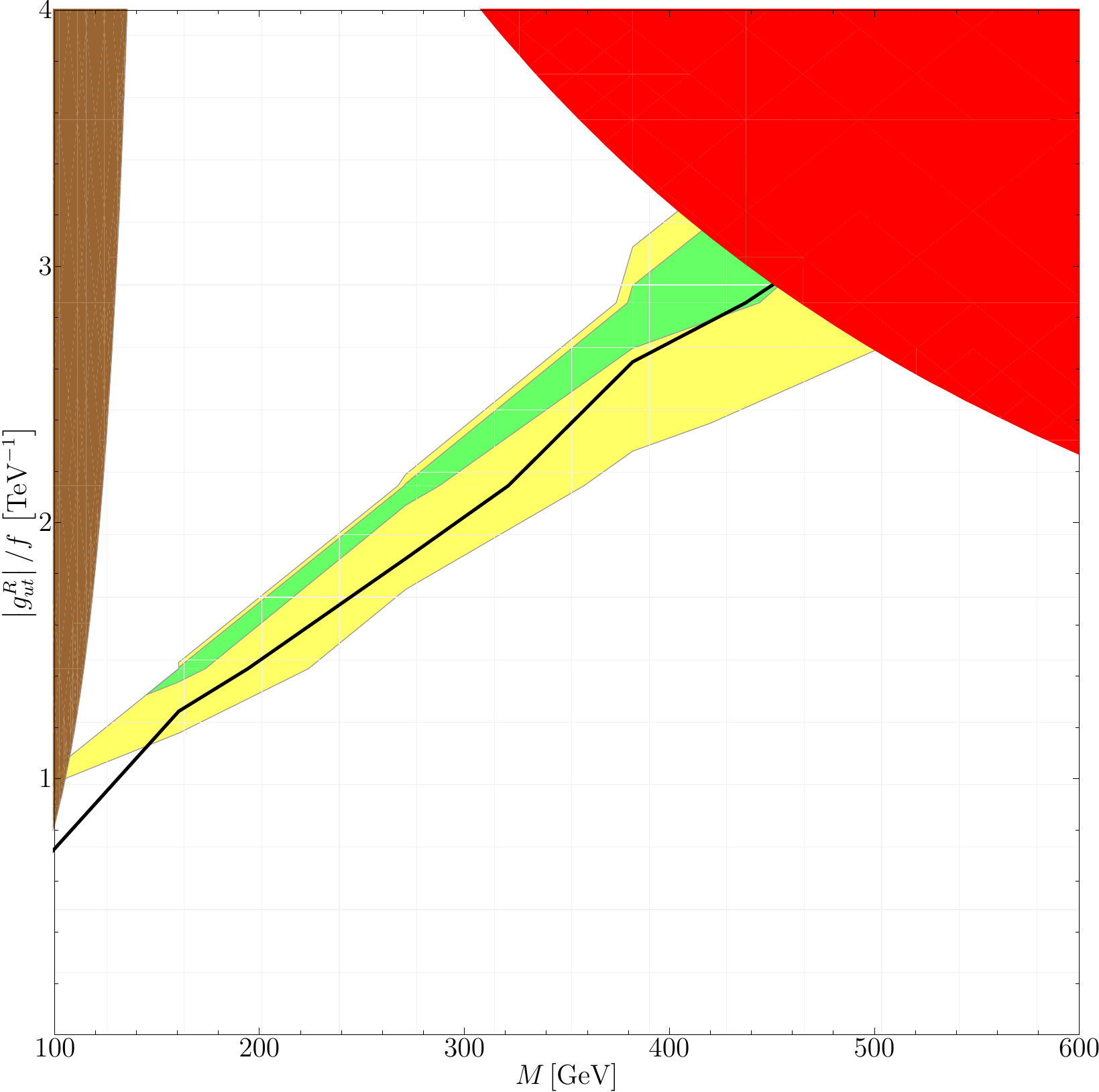}}
 \subfloat[$g^R_{uu} = g^R_{tt} = g^R_{ut} / 10 = 3 g^R_{tu}$]{\label{fig:chi4}\includegraphics[width=0.5\textwidth]{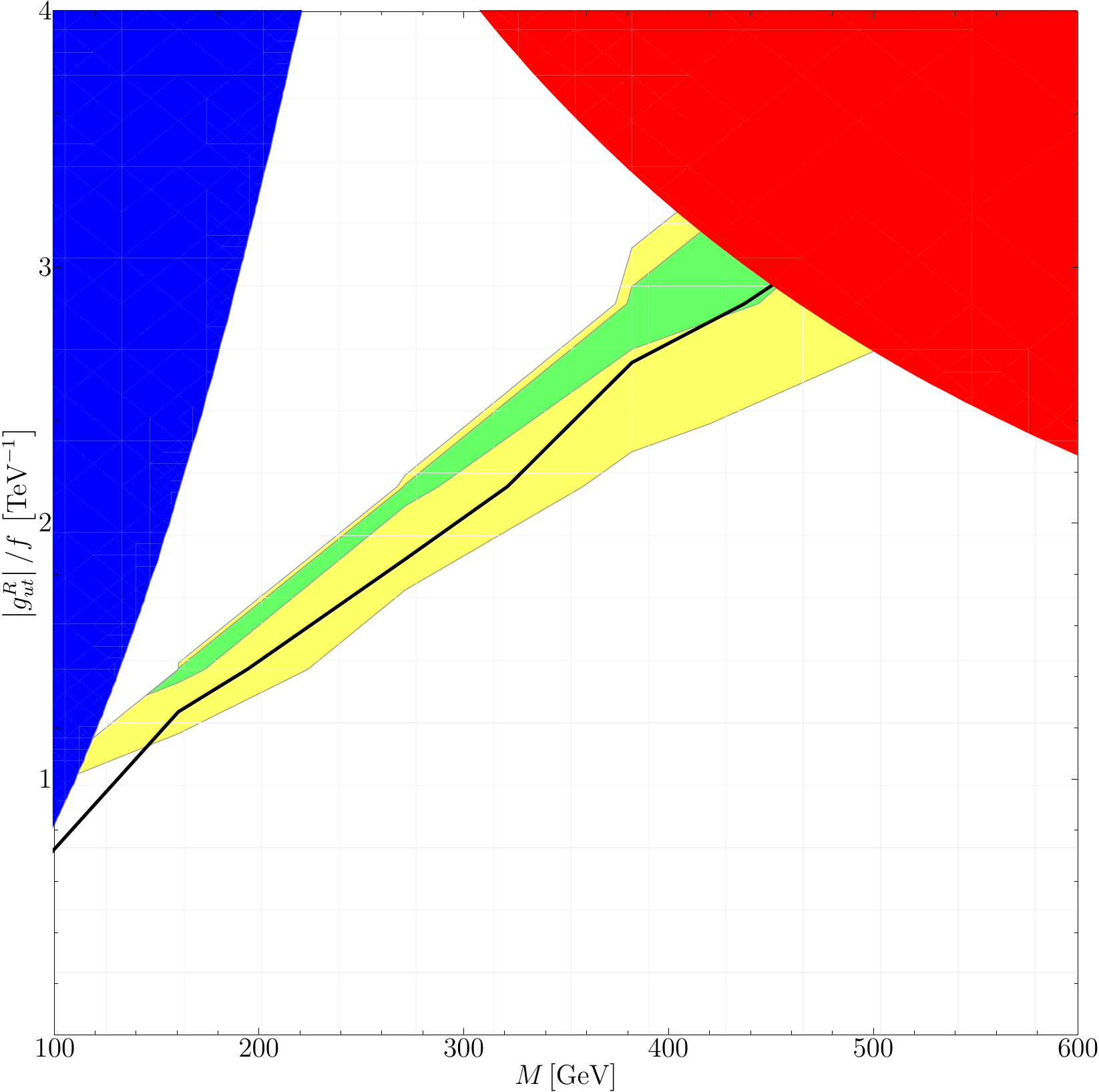}} \\
 \subfloat[$g^R_{uu} = g^R_{tt} = g^R_{ut} / 2 = 15 g^R_{tu}$]{\label{fig:chi3}\includegraphics[width=0.5\textwidth]{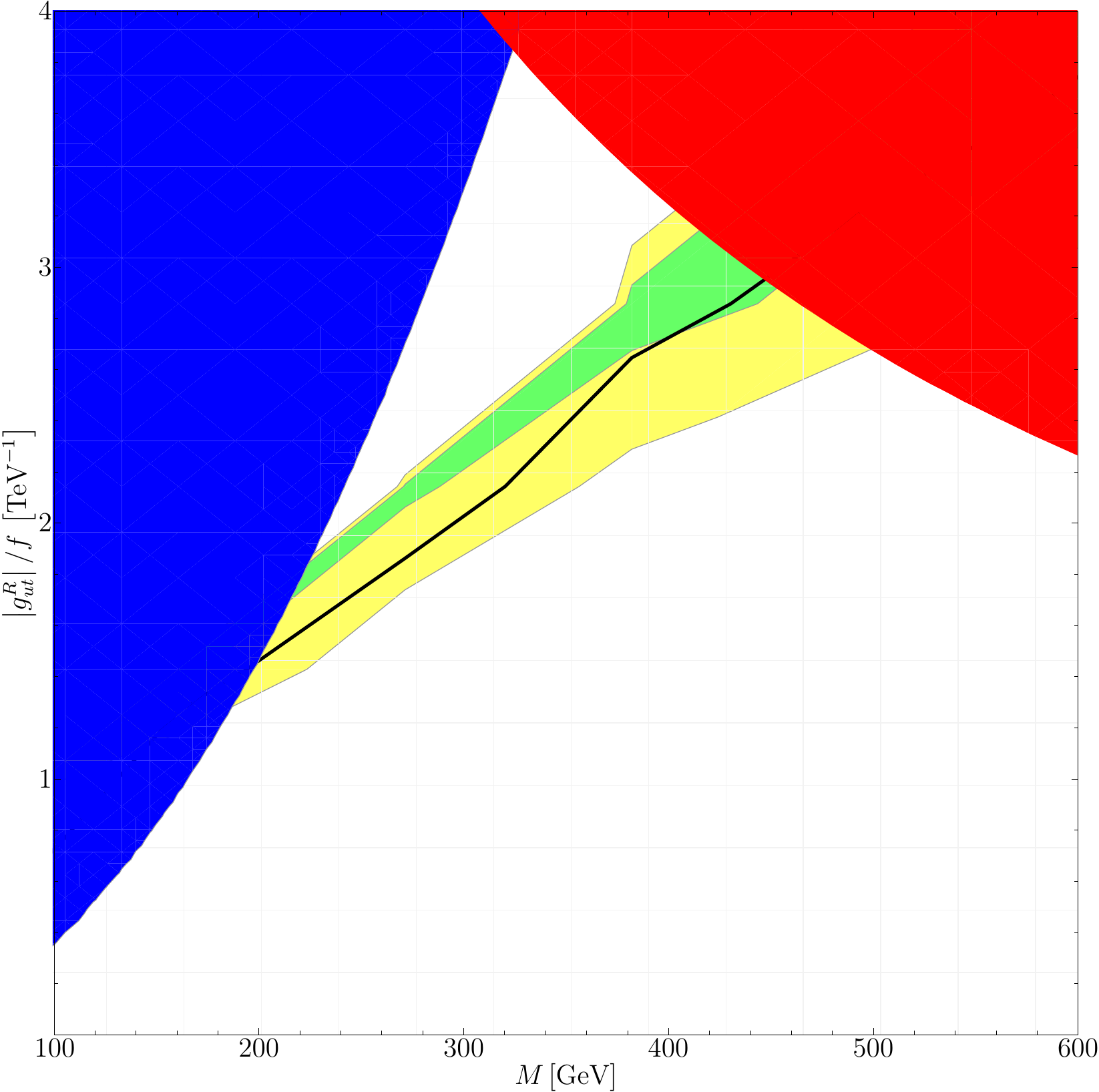}}
 \subfloat[$g^R_{uu} = g^R_{tt} = g^R_{ut} = 30 g^R_{tu}$]{\label{fig:chi2}\includegraphics[width=0.5\textwidth]{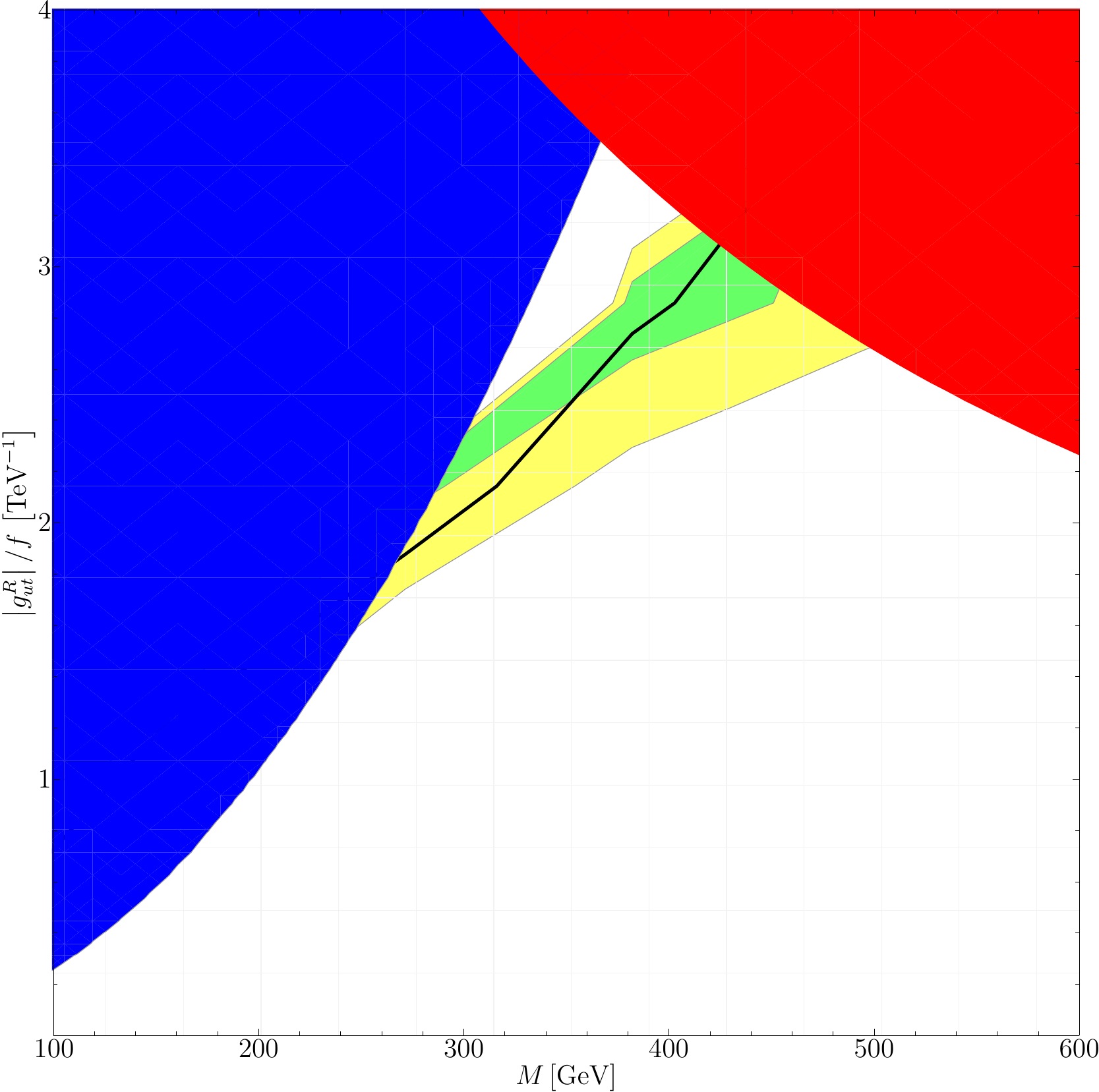}}
   \caption{Results of a global fit of the spin-2 model to Tevatron observables.  $A^{t\bar{t}}_{\text{high}}= 47.5\%$ is shown in black.  The 1 and 2$\sigma$ confidence regions of allowed parameters are shown in green and yellow respectively.  The blue, red, and brown regions are disfavored by constraints from same-sign top, EWPD, and the width of the top respectively.}
    \captionsetup{justification=raggedright}
 \label{fig:chi}
   \end{figure}

A global fit of the spin-2 model to the CDF measurements $A^{t\bar{t}}_{\text{high}},\, A^{t\bar{t}}_{\text{low}},\, \text{and}\,\sigma_{t\bar{t}}$ was performed using the method of least squares assuming the measurements are uncorrelated\footnote{Of course, the measurements actually are correlated, but this should not affect the conclusions we are able to draw from the fit in any qualitative way.}.  The scale $f$ was fixed to be 1 TeV and $g^R_{ut},\, g^R_{uu},\, \text{and}\,g^R_{tt}$ were left as free parameters for a given $M$.  The results of the fit are shown in Fig.~\ref{fig:chi}.  The 1 and 2$\sigma$ confidence regions of allowed parameter space are shown in green and yellow respectively.  The black line corresponds to $A^{t\bar{t}}_{\text{high}}= 47.5\%$.  This is not necessarily the best-fit value.  Experimentally disallowed parameter space due to the constraints from same-sign top production at the LHC, EWPD, and the width of the top quark are shown in blue, red, and brown respectively (see Section~\ref{sec:constraints} for a detailed discussion of experimental constraints on the model).  The spin-2 model is able to hit the central value of the forward-backward asymmetry in the high mass bin in a large region of the parameter space.

\begin{figure}
  \centering
    \subfloat[$A^{t\bar{t}}_{FB}$]{\label{fig:asym350}\includegraphics[width=0.5\textwidth]{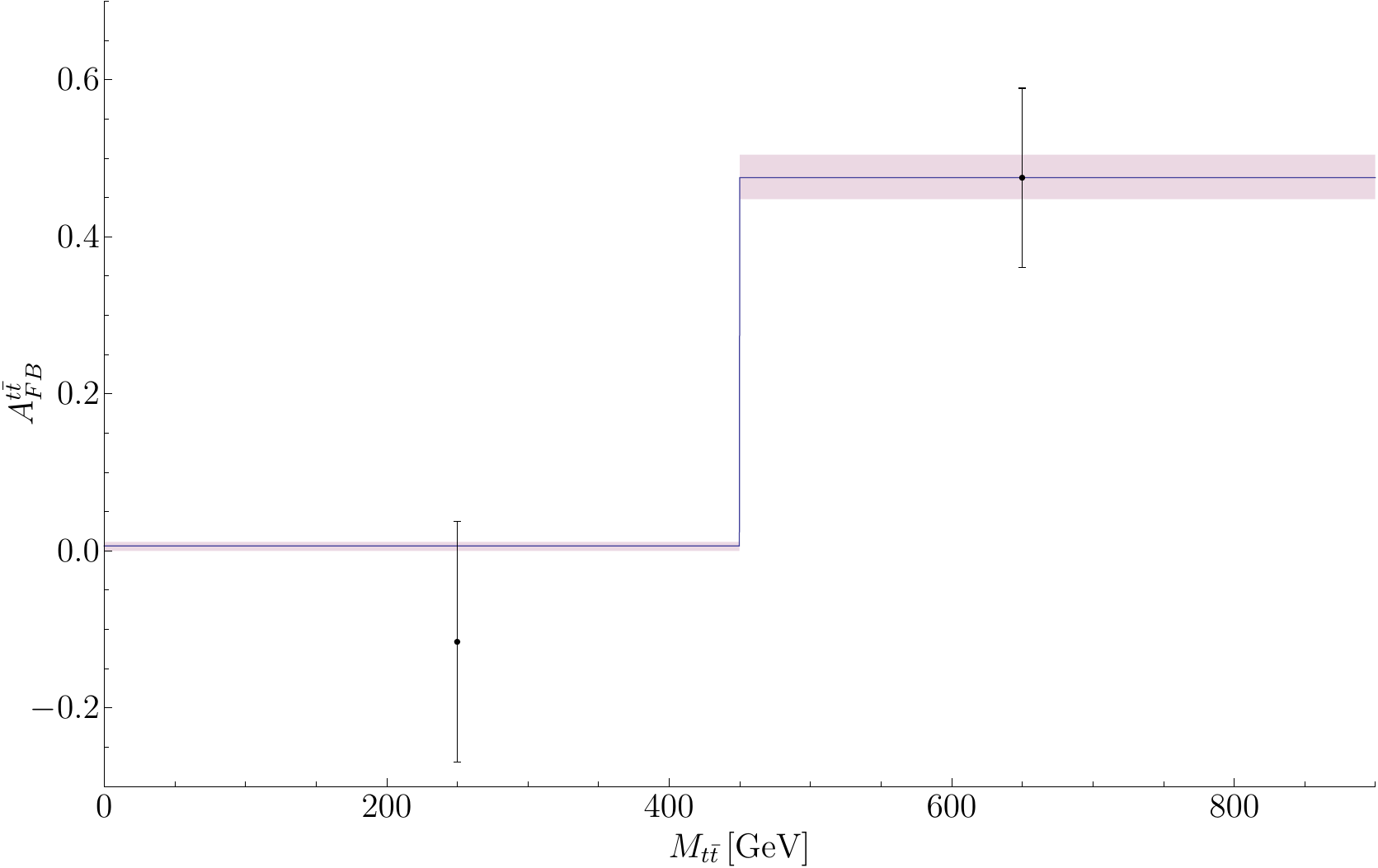}}
    \subfloat[$d\sigma_{t\bar{t}}/dM_{t\bar{t}}$]{\label{fig:diff350}\includegraphics[width=0.5\textwidth]{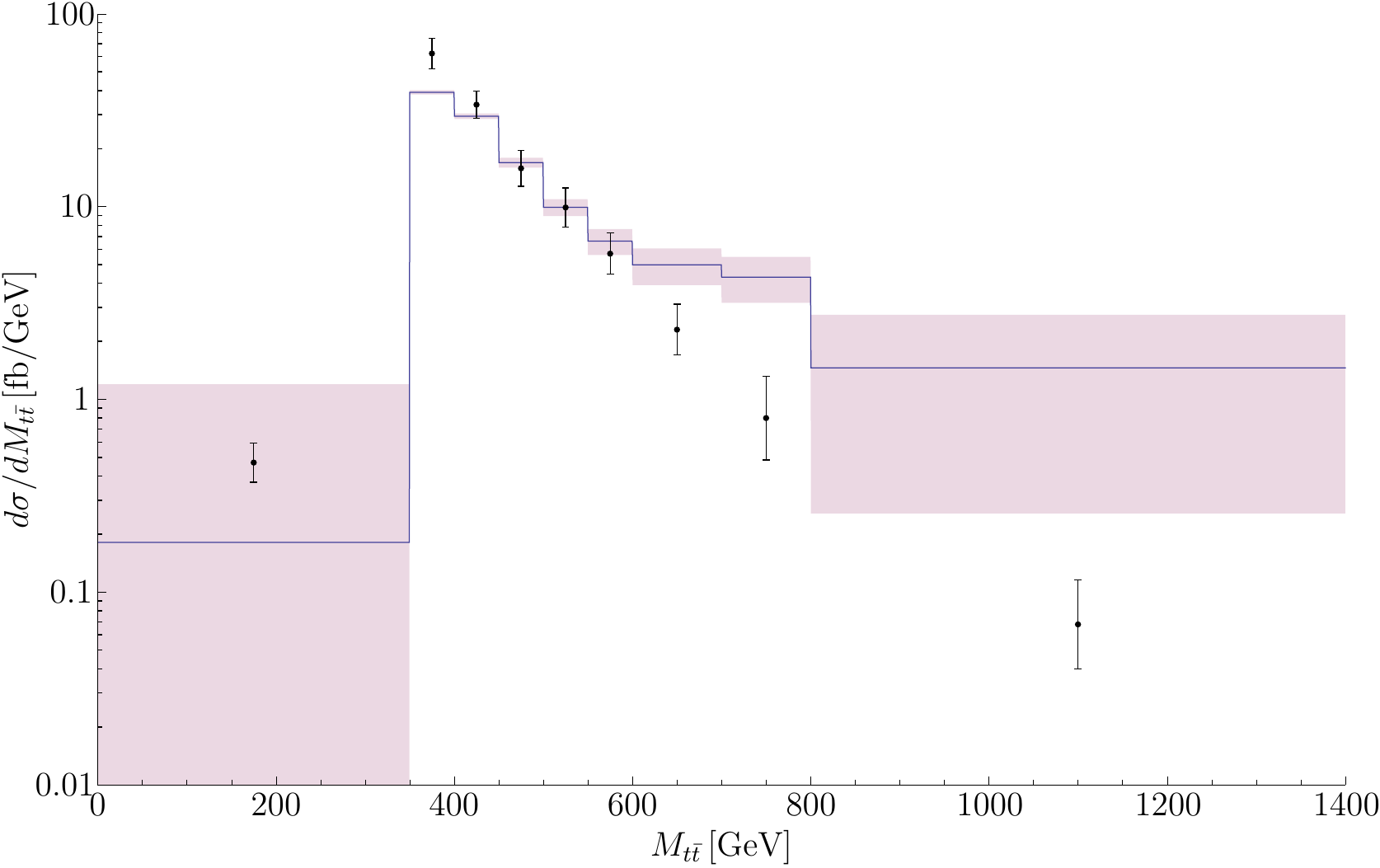}} 
    \caption{Prediction from the spin-2 model for $A^{t\bar{t}}_{FB}$ and $d\sigma_{t\bar{t}}/dM_{t\bar{t}}$ with $M = 350$ GeV.  The purple band represents the theoretical uncertainty from varying the factorization scale in the range $\mu = \{m_t / 2,\,2 m_t\}$.  This example hits the central value of $A^{t\bar{t}}_{FB}$ in the high bin and is within $1\sigma$ of the central value in the low bin.  Detector acceptance effects and the known increase in the measured value for $\sigma_{t\bar{t}}$ could account for the disagreement in the high mass bins for $d\sigma_{t\bar{t}}/dM_{t\bar{t}}$.}
 \label{fig:4}
   \end{figure}

Fig.~\ref{fig:asym350} shows the binned asymmetry predicted by the spin-2 model for $M = 350$ GeV, $\left|g^R_{ut}\right| / f = 2.36 /$TeV, and all other couplings set to zero.  The CDF measurements with error bars are also shown.  The purple band represents the theoretical uncertainty from varying the factorization scale in the range $\mu = \{m_t / 2,\,2 m_t\}$.  This combination of parameters hits the central value of $A^{t\bar{t}}_{FB}$ in the high bin and is within $1\sigma$ of the central value in the low bin.  The sum of the SM LO prediction plus the contribution from the spin-2 model with the same parameters as those used in Fig.~\ref{fig:asym350} for the binned differential cross section, $d\sigma_{t\bar{t}}/dM_{t\bar{t}}$, is shown in Fig.~\ref{fig:diff350}.  Again, the purple band represents the uncertainty in the PDF factorization scale, and the CDF measurements, as reported in Ref.~\cite{Aaltonen:2009iz}, are also shown.  The high bin values do not agree with the CDF measurements.  However, we have not taken into account any detector acceptance effects.  The deconvolution to the parton level done by CDF assumes the SM.  As shown in Ref.~\cite{Grinstein:2011dz}, model-dependent acceptance effects can reduce the cross section by as much as a factor of $\sim 1/2$ in the high bins.  Furthermore, the total cross section reported in Ref.~\cite{Aaltonen:2009iz} is $\sigma_{t\bar{t}} = 6.9 \pm 1.0$ pb, which is lower than the most recent measurements from both the CDF~\cite{Aaltonen:2010hza} and D\O ~\cite{Abazov:2011mi} collaborations.  It is reasonable to assume that detector acceptance effects and the known increase in the measured value for $\sigma_{t\bar{t}}$ could account for the disagreement in the high mass bins for $d\sigma_{t\bar{t}}/dM_{t\bar{t}}$.

\subsection{LHC Measurements, Predictions, and Results}
There is no forward-backward asymmetry at the LHC because of its symmetric initial state, $pp$, as opposed to the $p\bar{p}$ initial state at the Tevatron.  However, the same underlying physics that leads to $A_{FB}$ results in a \textit{charge asymmetry} at the LHC, which we define as 
\begin{equation}
A^y_C = \frac{\sigma\left(\Delta y^2 > 0\right) - \sigma\left(\Delta y^2 < 0\right)}{\sigma\left(\Delta y^2 > 0\right) +\sigma\left(\Delta y^2 < 0\right)},
\end{equation}
where $\Delta y^2$ is the difference of the squares of rapidities of the top quark and anti-top quark, $\Delta y^2 = y_t^2 - y_{\bar{t}}^2$.  The CMS collaboration reports~\cite{Chatrchyan:2011hk} the charge asymmetry of $A^y_C = \left(-1.3^{+4.0}_{-4.2}\right)\%$, which is consistent with the SM prediction $A^y_C = \left(1.15 \pm 0.06\right)\%$~\cite{Kuhn:2011ri}.  The value of $A^y_C$ reported~\cite{ATLAS:2012an} by the ATLAS collaboration, $A^y_C = \left(-2.4 \pm 2.8\right)\%$, is consistent with the measurement of CMS.  

ATLAS recently measured~\cite{:2012kg} the top quark production cross section with $\sqrt{s} = 7$ TeV to be $\sigma_{t\bar{t}} = 176^{+17}_{-14}$ pb, which is consistent with the CMS observation $\sigma_{t\bar{t}} = \left(154 \pm 18\right)$ pb~\cite{Chatrchyan:2011yy}.  A QCD prediction~\cite{Ahrens:2011mw} at approximate NNLO (using 1PI$_{\text{SCET}}$) yielded $\sigma_{t\bar{t}} = 155^{+11}_{-12}$ pb.  The EWS correction factor~\cite{Manohar:2012rs}, $\mathcal{R}_t = 0.98$, is used to compute the full SM prediction, $\sigma_{t\bar{t}} = 152^{+11}_{-12}$ pb.

\begin{figure}
  \centering
\includegraphics[width=0.7\textwidth]{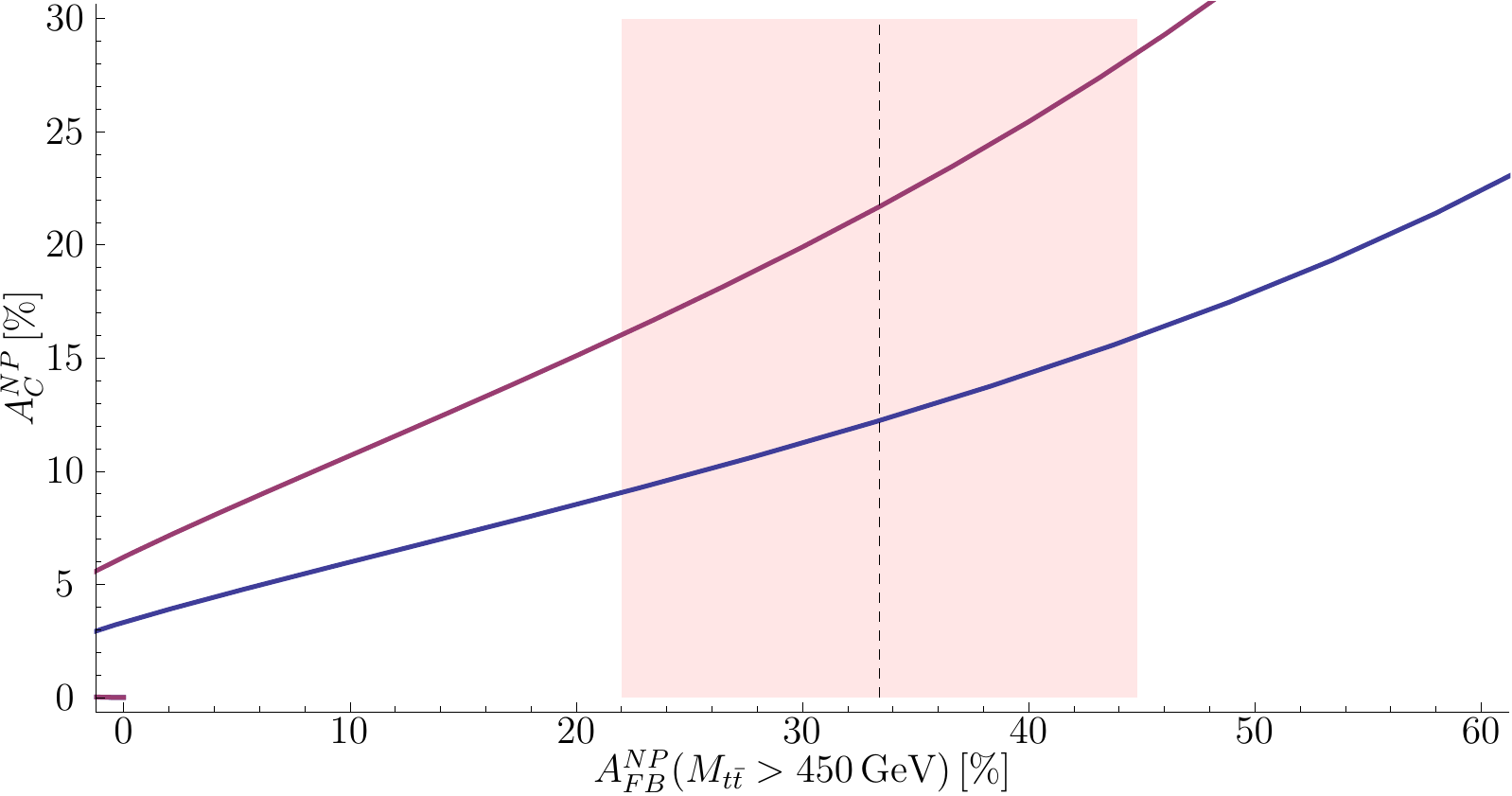}
    \caption{NP contributions to the inclusive charge asymmetry at the LHC and the forward-backward asymmetry at the Tevatron in the high mass bin.  The pink band is the $1\sigma$ error of the measurement $A^{t\bar{t}}_{\text{high}}$, and the dashed line is the difference between the measured value of $A^{t\bar{t}}_{\text{high}}$ and the corresponding SM prediction.  The blue and red curves are predictions of the spin-2 model for $M =100,\,200$ GeV respectively.}
 \label{fig:asymC}
   \end{figure}
Fig.~\ref{fig:asymC} shows $A^{y}_{C}$ at the LHC as a function of $A^{t\bar{t}}_{FB}$ at the Tevatron for the spin-2 model at hand.  The pink band is the $1\sigma$ error of the measurement $A^{t\bar{t}}_{\text{high}}$, with the central value given by the dashed line.  The blue and red curves are predictions of the spin-2 model for $M =100,\,200$ GeV respectively.  These predictions are not within $1\sigma$ of both measurements simultaneously.  This is a generic feature of any model that attempts to explain $A^{t\bar{t}}_{FB}$~\cite{AguilarSaavedra:2011hz,AguilarSaavedra:2011ug}.

If the only NP is a single spin-2 field, then the cutoff of the effective theory should be at least as large as the center-of-mass energy of the experiment.  The most optimistic estimate of the cutoff is $\Lambda \approx 4\pi\bar{f}$.  As argued in section~\ref{sec:spin2}, the cutoff is likely to be smaller than this estimate.  Fixing $f$ to be 1 TeV, $g_{ut}$ should be less than $\lsim 1.8$ if the effective theory is to be valid up to 7 TeV, as opposed to $g_{ut} \lsim 6.4$ for $\sqrt{s} = 1.96$ TeV at the Tevatron.  Adding heavier fields to the effective theory could raise the cutoff, as well as lead to a qualitatively different relation between the forward-backward and charge asymmetries measured at the Tevatron and LHC. Preliminary analysis has shown that interference between the virtual exchange of a lighter spin-2 particle and a heavier spin-2 state
could ameliorate the tension between the measurements of $A^{y}_{C}$ and $A^{t\bar{t}}_{FB}$. 
However, a more comprehensive study of this effect is needed for quantitative predictions, and we leave this for a future study. 

\begin{figure}
  \centering
  \includegraphics[width=0.7\textwidth]{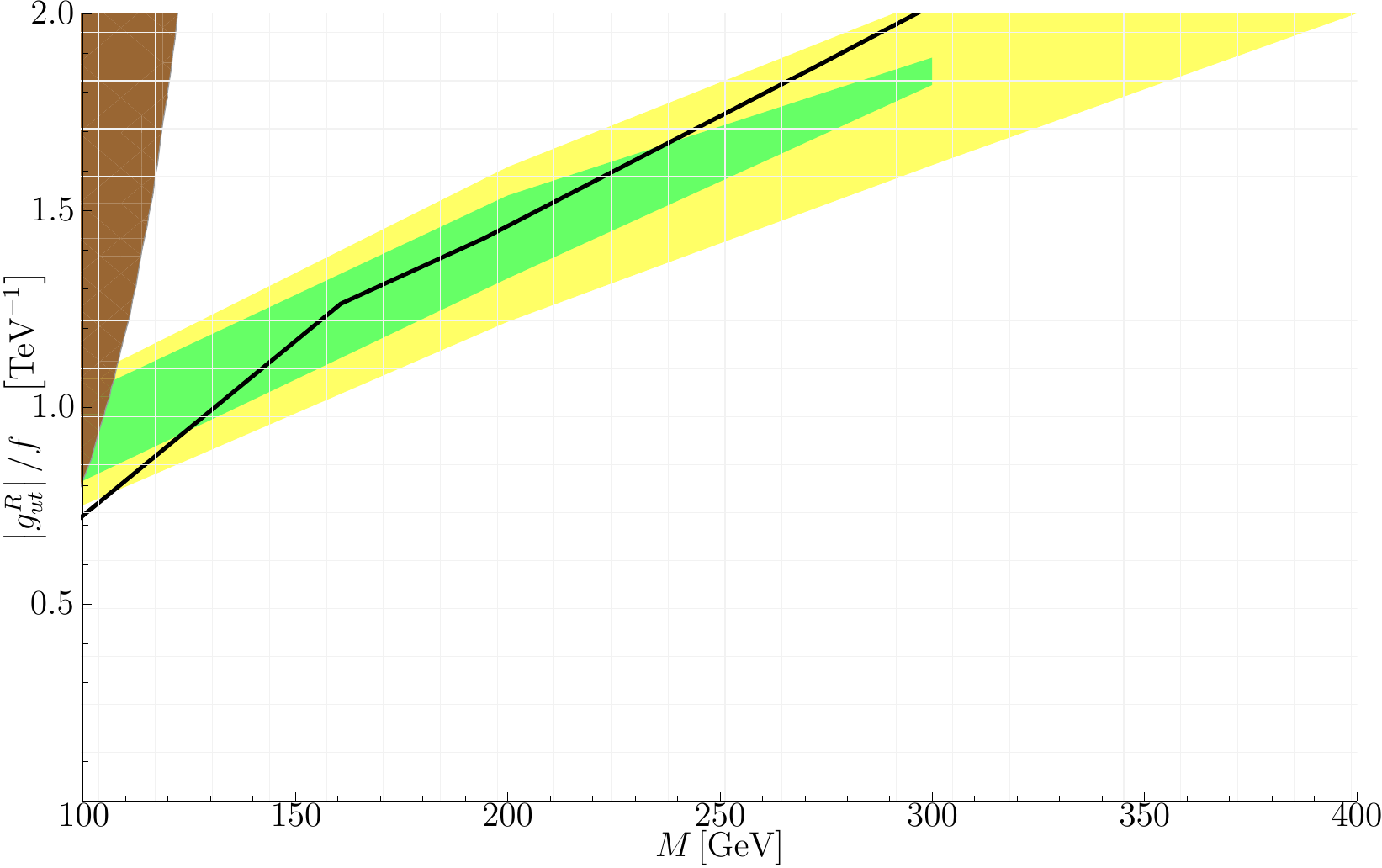}
    \caption{Fit to  $A^{t\bar{t}}_{\text{high}}$ (CDF) and $\sigma_{t\bar{t}}$ (ATLAS).  $f = 1$ TeV and $g^R_{ut}$ is a free parameter for a given $M$.  The 1 and 2$\sigma$ confidence regions of allowed parameter space are shown in green and yellow respectively.  The black line corresponds to $A^{t\bar{t}}_{\text{high}}= 47.5\%$.  Experimentally disallowed parameter space due to constraints the width of the top quark is shown in brown.}
 \label{fig:chiC}
   \end{figure}
A global fit of the spin-2 model to the CDF measurement $A^{t\bar{t}}_{\text{high}}$ and the ATLAS measurement $\sigma_{t\bar{t}}$ was performed using the method of least squares.  The scale $f$ was fixed to be 1 TeV and $g^R_{ut}$ was left as a free parameter for a given $M$.  Again, $g_{ut}$ should be less than $\lsim 1.8$ if effective theory is to be valid up to 7 TeV.  The results of the fit are shown in Fig.~\ref{fig:chiC}.  The 1 and 2$\sigma$ confidence regions of allowed parameter space are shown in green and yellow respectively.  The black line corresponds to $A^{t\bar{t}}_{\text{high}}= 47.5\%$.  This is not necessarily the best-fit value.  Experimentally disallowed parameter space due to constraints the width of the top quark is shown in brown.  As was the case with the fit to Tevatron observables, the spin-2 model is again able to hit the central value of the forward-backward asymmetry in the high mass bin in a large region of the parameter space.

There is an additional contribution to the $t\bar{t}$ production cross section at the LHC from single-top, spin-2 production where the spin-2 particle immediately decays into $u\bar{t}$.   Again, we approximate the $2\rightarrow3$ cross section as $\sigma(u\,g \rightarrow t\,\bar{t}\,u) \approx \sigma(u\,g \rightarrow t\,h_{\mu\nu})\times Br(h_{\mu\nu} \rightarrow u\,\bar{t})$.  We used the ATLAS collaboration's~\cite{:2012kg} cut, $|\eta| < 2.5$ when selecting muon and jet candidates associated with $t\bar{t}$ production, when we calculated the spin-2 contribution to this cross section.   For $M = 200$ GeV and $f = 1$ TeV, requiring $\sigma_{t\bar{t}}$ to be within 1$\sigma$ of the measured value limits $|g_{ut}|$ to be less than 1.55 assuming $Br(h_{\mu\nu} \rightarrow u\bar{t}) = 1$.  Alternatively, one may allow $|g_{ut}|$ to reach its maximum allowed value in the effective theory of approximately 1.8 by requiring $Br(h_{\mu\nu} \rightarrow u\bar{t}) \lsim 0.9$  The contribution to $\sigma_{t\bar{t}}$ from this channel falls with the mass of the spin-2 particle such that theoretical considerations quickly become the dominant constraint on $|g_{ut}|$.

\subsection{Comments on Differential Measurements}
At the time this paper was submitted for publication, new measurements of the charge asymmetry at the LHC at the differential level were reported by the ATLAS~\cite{ATLAS:2012an} and CMS~\cite{CMS-PAS-TOP-11-030} collaborations.  CMS also recently measured~\cite{CMS-PAS-TOP-11-013} the normalized, differential $t\bar{t}$ production cross section.  By normalizing the differential cross section to the total cross section, certain systematic uncertainties and all  normalization uncertainties cancel out, leading to a particularly precise measurement.    In addition, the CDF collaboration updated~\cite{CDF-Note-10807} its analysis of the forward-backward asymmetry at the Tevatron to include the full Run II dataset.  It is observed that $A^{t\bar{t}}_{FB}$ has an approximately linear dependence on both $M_{t\bar{t}}$ and $\Delta y$.  In this section, we discuss the spin-2 model's predictions for these differential measurements.

\begin{figure}
  \centering
   \subfloat[$A^{t\bar{t}}_{FB}$ vs. $M_{t\bar{t}}$]{\label{fig:cdfAsymUpdate}\includegraphics[width=0.49\textwidth]{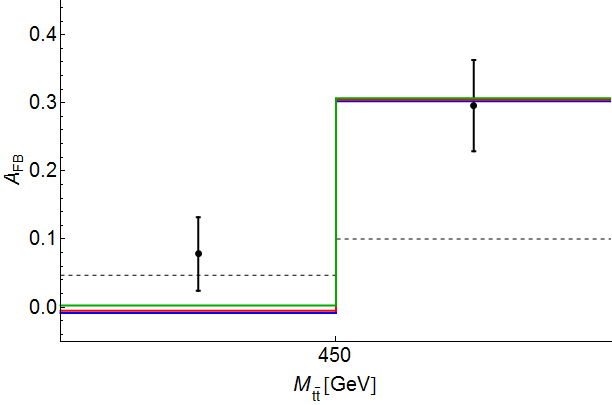}}
 \subfloat[$(d\sigma / dM_{t\bar{t}}) / \sigma$ vs. $M_{t\bar{t}}$]{\label{fig:cmsDiffXSection}\includegraphics[width=0.49\textwidth]{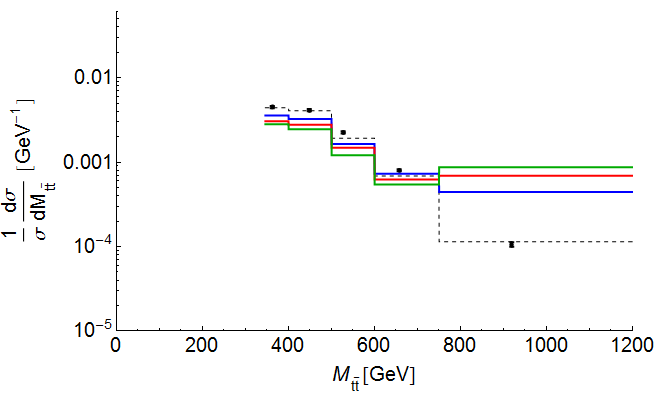}} \\
   \subfloat[$A^{t\bar{t}}_{C}$ vs. $M_{t\bar{t}}$, ATLAS data]{\label{fig:atlasDiffAsym}\includegraphics[width=0.49\textwidth]{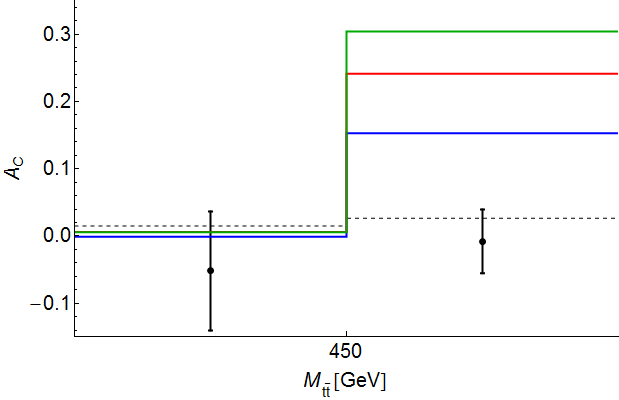}}
 \subfloat[$A^{t\bar{t}}_{C}$ vs. $M_{t\bar{t}}$, CMS data]{\label{fig:cmsDiffAsym}\includegraphics[width=0.49\textwidth]{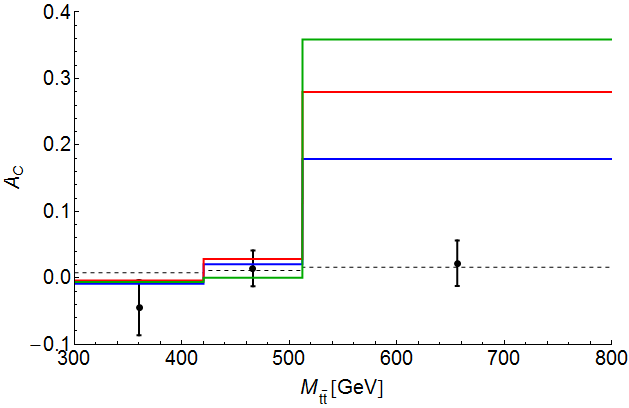}} 
   \caption{Predictions for forward-backward asymmetry at the Tevatron (upper left), normalized, differential cross section at the LHC (upper right), and charge asymmetry at the LHC (bottom).  The blue, red, and green lines correspond to a spin-2 mass and coupling, $g_{ut}/f$, of (100 GeV, 0.85 TeV$^{-1}$), (200 GeV, 1.38 TeV$^{-1}$), and (300 GeV, 1.84 TeV$^{-1}$) respectively.}
 \label{fig:update1}
   \end{figure}

Fig.~\ref{fig:update1} shows predictions of the spin-2 model for the forward-backward asymmetry at the Tevatron, the normalized differential top quark production cross section at the LHC, and the charge-asymmetry at the LHC.  The blue, red, and green lines correspond to a spin-2 mass of $\{100,\,200,\,300\}$ GeV and a coupling $g_{ut}/f = \{0.85,\,1.38,1.84\}\,\text{TeV}^{-1}$.  The dashed lines are the SM values for these observables.  These calculations were made using \texttt{FeynRules}~\cite{Christensen:2008py} interfaced with \texttt{MadGraph 5}~\cite{Alwall:2011uj}.  

As was the case for the inclusive charge asymmetry, these predictions are not simultaneously within $1\sigma$ of both the differential charge asymmetry and the forward-backward asymmetry measurements.  To the best of our knowledge, this is a generic feature of any model for $A^{t\bar{t}}_{FB}$ proposed before this paper was submitted for publication, see for example Refs.~\cite{AguilarSaavedra:2011hz,AguilarSaavedra:2011ug}.  After this paper was submitted for publication, Drobnak \textit{et al}. discovered~\cite{Drobnak:2012cz} a class of models that can accommodate both measurements simultaneously.  Based on the results of~\cite{Drobnak:2012cz}, a spin-2 field that is charged under certain representations of the SM gauge group may produce a large $A^{t\bar{t}}_{FB}$ and a negligible $A^{t\bar{t}}_{C}$.  Verifying this claim is beyond the scope of this work, but it will be investigated in a future project.

The shape of the normalized, differential cross section predicted by the spin-2 model does not agree with the CMS measurement.  However, as was the case for the differential cross section at the Tevatron, we have not taken into account any detector acceptance effects.  As shown in Ref.~\cite{Grinstein:2011dz}, model-dependent acceptance effects can reduce the cross section by as much as a factor of $\sim 1/2$ in the high bins.    No uncertainties from the choice of scale or PDFs have been included either.  It is reasonable to assume that these unaccounted for effects could help to ameliorate some of the tension between the measured and predicted value of $(d\sigma/dM_{t\bar{t}})/\sigma$.  Nevertheless, the shape of differential cross section constrains the spin-2 model's parameter space.  The question becomes, how large of an asymmetry can generated once this constraint is taken into account.

\section{Conclusions}
\label{sec:con}
If it persists, the anomalously large top quark forward-backward asymmetry observed at the Tevatron is an  indication of physics beyond the Standard Model. Since its appearance, many models involving new scalar, as well as vector particles around the weak scale have been proposed to address the anomaly. However, it has proven to be hard to raise the theoretical prediction to the central value of the CDF measurement in the high mass bin.  We have shown that there is parameter space in this model, consistent with various experimental constraints that could accommodate the CDF measurement of $A_{FB}^{t\bar{t}}(M_{t\bar{t}}>450\text{GeV})$ of $47.5\%$.  The peculiar derivative coupling of a spin-2 particle to fermions naturally leads to strong sensitivity of the asymmetry to the $t\bar t$ invariant mass. As a result, the picture of the top asymmetry increasing with energy observed by CDF naturally fits in this framework.  If the observed $A_{FB}^{t\bar{t}}$ holds, it would be interesting to study the experimental bounds as well as the phenomenology of this model in more detail.


\begin{acknowledgements}
We thank Aneesh Manohar, Mike Trott, and David C. Stone for valuable discussions. This work has been supported in part by the U.S. Department of Energy under contract No. DOE-FG03-97ER40546.
\end{acknowledgements}

\appendix
\section{Form Factors} \label{sec:ap}
The combinations of coefficients that appear in Eqs.~\eqref{eq:T},~\eqref{eq:Tint},~\eqref{eq:S} and~\eqref{eq:ST} are:
\begin{align*}
C_1 &= \left|g^R_{ut}\right|^2+ \left|g^R_{tu}\right|^2 - \left|g^L_{ut}\right|^2-\left|g^L_{tu}\right|^2, \displaybreak[0]\\
C_2 &= \left|g^R_{ut}\right|^2 + \left|g^R_{tu}\right|^2+ \left|g^L_{ut}\right|^2+\left|g^L_{tu}\right|^2, \displaybreak[0]\\
C_3 &= \left(\left|g^R_{uu}\right|^2 + \left|g^L_{uu}\right|^2\right)\left(\left|g^R_{tt}\right|^2 + \left|g^L_{tt}\right|^2\right),\displaybreak[0] \\
C_4 &= \left(\left|g^R_{uu}\right|^2 + \left|g^L_{uu}\right|^2\right)g^R_{tt}\left(g^L_{tt}\right)^{\ast} +\text{c.c.}, \displaybreak[0]\\
C_5 &= \left(\left|g^R_{uu}\right|^2 - \left|g^L_{uu}\right|^2\right)\left(\left|g^R_{tt}\right|^2 - \left|g^L_{tt}\right|^2\right), \displaybreak[0]\\
C_6 &= \left(\left|g^R_{ut}\right|^2+\left|g^R_{tu}\right|^2\right)g^R_{uu}\left(g^R_{tt}\right)^{\ast} +\left( \left|g^L_{ut}\right|^2+\left|g^L_{tu}\right|^2\right)g^L_{uu}\left(g^L_{tt}\right)^{\ast} +\text{c.c.},\displaybreak[0] \\
C_7 &= \left(\left|g^R_{ut}\right|^2+\left|g^R_{tu}\right|^2\right)g^R_{uu}\left(g^L_{tt}\right)^{\ast} + \left(\left|g^L_{ut}\right|^2+\left|g^L_{tu}\right|^2\right)g^L_{uu}\left(g^R_{tt}\right)^{\ast} +\text{c.c.}.
\end{align*}

The form factors in Eqs.~\eqref{eq:T},~\eqref{eq:Tint},~\eqref{eq:S} and~\eqref{eq:ST} are:
\begin{align*}
F_{1}(x,\,y) &= \left(2y-2-x\right)\left(2y^3+4xy(4+x)-y^2(6+5x)-x(8+8x+x^2)\right),\displaybreak[0] \\
F_{2}(x,\,y) &= -y\left[ \left(2y^3+4xy(4+x)-y^2(6+5x)-x(8+8x+x^2)\right)\right. \\
&\quad\left. + \left(2y-2-x\right)\left(y^2-2xy+x(2+x)\right)\right],\displaybreak[0] \\
F_{3}(x,\,y) &= y^2\left[y^2-2xy+x(2+x)\right], \displaybreak[0]\\
F_{4}(x,\,y) &= 18\left(32 +64x +42x^2 + 10x^3 + x^4\right)-36y\left(64+82x+28x^2+3x^3\right) \\
&\quad+y^2\left(2712 +1764x +235x^2\right) -2y^3(492 +109x) +73y^4, \displaybreak[0]\\
F_{5}(x,\,y) &= -2y\left[-3\left(192+276x+104x^2+7x^3\right) + 2y\left(672 +480x +43x^2\right)\right. \\ 
&\quad\left. -y^2(720 +109x) +44y^3\right],\displaybreak[0]\\
F_{6}(x,\,y) &= 3\left[3x^2(2+x)^2-2xy\left(48+52x+11x^2\right) + y^2\left(256+332x+65x^2\right)\right. \\
 &\quad\left. -4y^3(72+19x) +30y^4\right],\displaybreak[0]\\
F_{7}(x,\,y) &= -8y\left[-3x^2(2+x) +2xy(12+7x) -y^2(24+19x) +8y^3\right],\displaybreak[0]\\
F_{8}(x,\,y) &= 16y^2(x-y)^2,\displaybreak[0]\\
F_{9}(x,\,y) &= 6(1+x)^2(4+x) - y\left(48+78x+23x^2\right) +y^2(37+28x) -11y^3,\displaybreak[0] \\
F_{10}(x,\,y) &= 3x^2(2+x) +y\left(24+3x-8x^2\right) + y^2(7x-8) -2y^3,\displaybreak[0] \\
F_{11}(x,\,y) &= 4y\left[x^2 + y(1-2x) +y^2\right],\displaybreak[0] \\
F_{12}(x,\,y) &= 1+ 10x +42x^2 +64x^3 + 32x^4 - 2y\left(3+28x +82x^2 +64x^3\right) \\
&\quad+ 2y^2\left(9+68x+96x^2\right) -4y^3(9+32x) +32y^4,\displaybreak[0] \\
F_{13}(x,\,y) &= 2y\left(1+2x+2x^2-4y(1+x)+2y^2\right), \displaybreak[0]\\
F_{14}(x,\,y) &= (1+2x-2y)\left(1+8x+8x^2-4y(1+4x)+8y^2\right),\displaybreak[0] \\
F_{15}(x,\,y) &= 6(1+x)^2\left(4+17x+4x^2\right) -24y\left(5+22x+22x^2+5x^3\right)\\
&\quad+y^2\left(253+596x+216x^2\right) -2y^3(109+84x) + 48y^4,\displaybreak[0] \\
F_{16}(x,\,y) &= y\left[3\left(8+33x+32x^2+8x^3\right) -2y\left(40+95x+36x^2\right)+y^2(94+72x)-24y^3\right], \displaybreak[0]\\
F_{17}(x,\,y) &= 4y^2(1+2x-2y),\displaybreak[0] \\
F_{18}(x,\,y) &= y\left[24+78x+51x^2+4x^3-4y\left(18+28x+3x^2\right)+y^2(61+12x)-4y^3\right],\displaybreak[0] \\
F_{19}(x,\,y) &= 3x^2\left(6+11x+4x^2\right) -4xy\left(12+36x+17x^2\right)+y^2\left(24+191x+132x^2\right) \\
&\quad-4y^3(20+27x)+32y^4, \displaybreak[0]\\
F_{20}(x,\,y) &= 4y(x-y)\left(x(3+4x)-y(1+8x)+4y^2\right).
\end{align*}

\bibliography{spin2v5}

\begin{thebibliography}{65}
\expandafter\ifx\csname natexlab\endcsname\relax\def\natexlab#1{#1}\fi
\expandafter\ifx\csname bibnamefont\endcsname\relax
  \def\bibnamefont#1{#1}\fi
\expandafter\ifx\csname bibfnamefont\endcsname\relax
  \def\bibfnamefont#1{#1}\fi
\expandafter\ifx\csname citenamefont\endcsname\relax
  \def\citenamefont#1{#1}\fi
\expandafter\ifx\csname url\endcsname\relax
  \def\url#1{\texttt{#1}}\fi
\expandafter\ifx\csname urlprefix\endcsname\relax\def\urlprefix{URL }\fi
\providecommand{\bibinfo}[2]{#2}
\providecommand{\eprint}[2][]{\url{#2}}

\bibitem[{\citenamefont{{CDF
  Collaboration}}(2011{\natexlab{a}})}]{CDFnote:10584}
\bibinfo{author}{\bibnamefont{{CDF Collaboration}}}
  (\bibinfo{year}{2011}{\natexlab{a}}), \eprint{CDF Note 10584}.

\bibitem[{\citenamefont{{D0
  Collaboration}}(2011{\natexlab{a}})}]{Abazov:2011rq}
\bibinfo{author}{\bibnamefont{{D0 Collaboration}}},
  \bibinfo{journal}{Phys.Rev.} \textbf{\bibinfo{volume}{D84}},
  \bibinfo{pages}{112005} (\bibinfo{year}{2011}{\natexlab{a}}),
  \eprint{1107.4995}.

\bibitem[{\citenamefont{Kamenik et~al.}(2011)\citenamefont{Kamenik, Shu, and
  Zupan}}]{Kamenik:2011wt}
\bibinfo{author}{\bibfnamefont{J.~F.} \bibnamefont{Kamenik}},
  \bibinfo{author}{\bibfnamefont{J.}~\bibnamefont{Shu}}, \bibnamefont{and}
  \bibinfo{author}{\bibfnamefont{J.}~\bibnamefont{Zupan}}
  (\bibinfo{year}{2011}), \eprint{1107.5257}.

\bibitem[{\citenamefont{{CDF
  Collaboration}}(2011{\natexlab{b}})}]{Aaltonen:2011kc}
\bibinfo{author}{\bibnamefont{{CDF Collaboration}}},
  \bibinfo{journal}{Phys.Rev.} \textbf{\bibinfo{volume}{D83}},
  \bibinfo{pages}{112003} (\bibinfo{year}{2011}{\natexlab{b}}),
  \eprint{1101.0034}.

\bibitem[{\citenamefont{Shu et~al.}(2010)\citenamefont{Shu, Tait, and
  Wang}}]{Shu:2009xf}
\bibinfo{author}{\bibfnamefont{J.}~\bibnamefont{Shu}},
  \bibinfo{author}{\bibfnamefont{T.~M.} \bibnamefont{Tait}}, \bibnamefont{and}
  \bibinfo{author}{\bibfnamefont{K.}~\bibnamefont{Wang}},
  \bibinfo{journal}{Phys.Rev.} \textbf{\bibinfo{volume}{D81}},
  \bibinfo{pages}{034012} (\bibinfo{year}{2010}), \eprint{0911.3237}.

\bibitem[{\citenamefont{Dorsner et~al.}(2010)\citenamefont{Dorsner, Fajfer,
  Kamenik, and Kosnik}}]{Dorsner:2009mq}
\bibinfo{author}{\bibfnamefont{I.}~\bibnamefont{Dorsner}},
  \bibinfo{author}{\bibfnamefont{S.}~\bibnamefont{Fajfer}},
  \bibinfo{author}{\bibfnamefont{J.~F.} \bibnamefont{Kamenik}},
  \bibnamefont{and} \bibinfo{author}{\bibfnamefont{N.}~\bibnamefont{Kosnik}},
  \bibinfo{journal}{Phys.Rev.} \textbf{\bibinfo{volume}{D81}},
  \bibinfo{pages}{055009} (\bibinfo{year}{2010}), \eprint{0912.0972}.

\bibitem[{\citenamefont{Dorsner et~al.}(2011)\citenamefont{Dorsner, Drobnak,
  Fajfer, Kamenik, and Kosnik}}]{Dorsner:2011ai}
\bibinfo{author}{\bibfnamefont{I.}~\bibnamefont{Dorsner}},
  \bibinfo{author}{\bibfnamefont{J.}~\bibnamefont{Drobnak}},
  \bibinfo{author}{\bibfnamefont{S.}~\bibnamefont{Fajfer}},
  \bibinfo{author}{\bibfnamefont{J.~F.} \bibnamefont{Kamenik}},
  \bibnamefont{and} \bibinfo{author}{\bibfnamefont{N.}~\bibnamefont{Kosnik}}
  (\bibinfo{year}{2011}), \eprint{1107.5393}.

\bibitem[{\citenamefont{Nelson et~al.}(2011)\citenamefont{Nelson, Okui, and
  Roy}}]{Nelson:2011us}
\bibinfo{author}{\bibfnamefont{A.~E.} \bibnamefont{Nelson}},
  \bibinfo{author}{\bibfnamefont{T.}~\bibnamefont{Okui}}, \bibnamefont{and}
  \bibinfo{author}{\bibfnamefont{T.~S.} \bibnamefont{Roy}}
  (\bibinfo{year}{2011}), \eprint{1104.2030}.

\bibitem[{\citenamefont{Cheung et~al.}(2009)\citenamefont{Cheung, Keung, and
  Yuan}}]{Cheung:2009ch}
\bibinfo{author}{\bibfnamefont{K.}~\bibnamefont{Cheung}},
  \bibinfo{author}{\bibfnamefont{W.-Y.} \bibnamefont{Keung}}, \bibnamefont{and}
  \bibinfo{author}{\bibfnamefont{T.-C.} \bibnamefont{Yuan}},
  \bibinfo{journal}{Phys.Lett.} \textbf{\bibinfo{volume}{B682}},
  \bibinfo{pages}{287} (\bibinfo{year}{2009}), \eprint{0908.2589}.

\bibitem[{\citenamefont{Patel and Sharma}(2011)}]{Patel:2011eh}
\bibinfo{author}{\bibfnamefont{K.~M.} \bibnamefont{Patel}} \bibnamefont{and}
  \bibinfo{author}{\bibfnamefont{P.}~\bibnamefont{Sharma}},
  \bibinfo{journal}{JHEP} \textbf{\bibinfo{volume}{1104}}, \bibinfo{pages}{085}
  (\bibinfo{year}{2011}), \eprint{1102.4736}.

\bibitem[{\citenamefont{Blum et~al.}(2011)\citenamefont{Blum, Hochberg, and
  Nir}}]{Blum:2011fa}
\bibinfo{author}{\bibfnamefont{K.}~\bibnamefont{Blum}},
  \bibinfo{author}{\bibfnamefont{Y.}~\bibnamefont{Hochberg}}, \bibnamefont{and}
  \bibinfo{author}{\bibfnamefont{Y.}~\bibnamefont{Nir}},
  \bibinfo{journal}{JHEP} \textbf{\bibinfo{volume}{1110}}, \bibinfo{pages}{124}
  (\bibinfo{year}{2011}), \eprint{1107.4350}.

\bibitem[{\citenamefont{Stone and Uttayarat}(2012)}]{Stone:2011dn}
\bibinfo{author}{\bibfnamefont{D.~C.} \bibnamefont{Stone}} \bibnamefont{and}
  \bibinfo{author}{\bibfnamefont{P.}~\bibnamefont{Uttayarat}},
  \bibinfo{journal}{JHEP} \textbf{\bibinfo{volume}{1201}}, \bibinfo{pages}{096}
  (\bibinfo{year}{2012}), \eprint{1111.2050}.

\bibitem[{\citenamefont{de~la Puente}(2012)}]{delaPuente:2011iu}
\bibinfo{author}{\bibfnamefont{A.}~\bibnamefont{de~la Puente}},
  \bibinfo{journal}{JHEP} \textbf{\bibinfo{volume}{1202}}, \bibinfo{pages}{016}
  (\bibinfo{year}{2012}), \eprint{1111.4488}.

\bibitem[{\citenamefont{Barger et~al.}(2010)\citenamefont{Barger, Keung, and
  Yu}}]{Barger:2010mw}
\bibinfo{author}{\bibfnamefont{V.}~\bibnamefont{Barger}},
  \bibinfo{author}{\bibfnamefont{W.-Y.} \bibnamefont{Keung}}, \bibnamefont{and}
  \bibinfo{author}{\bibfnamefont{C.-T.} \bibnamefont{Yu}},
  \bibinfo{journal}{Phys.Rev.} \textbf{\bibinfo{volume}{D81}},
  \bibinfo{pages}{113009} (\bibinfo{year}{2010}), \eprint{1002.1048}.

\bibitem[{\citenamefont{Shelton and Zurek}(2011)}]{Shelton:2011hq}
\bibinfo{author}{\bibfnamefont{J.}~\bibnamefont{Shelton}} \bibnamefont{and}
  \bibinfo{author}{\bibfnamefont{K.~M.} \bibnamefont{Zurek}},
  \bibinfo{journal}{Phys.Rev.} \textbf{\bibinfo{volume}{D83}},
  \bibinfo{pages}{091701} (\bibinfo{year}{2011}), \eprint{1101.5392}.

\bibitem[{\citenamefont{Grinstein
  et~al.}(2011{\natexlab{a}})\citenamefont{Grinstein, Kagan, Trott, and
  Zupan}}]{Grinstein:2011yv}
\bibinfo{author}{\bibfnamefont{B.}~\bibnamefont{Grinstein}},
  \bibinfo{author}{\bibfnamefont{A.~L.} \bibnamefont{Kagan}},
  \bibinfo{author}{\bibfnamefont{M.}~\bibnamefont{Trott}}, \bibnamefont{and}
  \bibinfo{author}{\bibfnamefont{J.}~\bibnamefont{Zupan}},
  \bibinfo{journal}{Phys.Rev.Lett.} \textbf{\bibinfo{volume}{107}},
  \bibinfo{pages}{012002} (\bibinfo{year}{2011}{\natexlab{a}}),
  \eprint{1102.3374}.

\bibitem[{\citenamefont{Grinstein
  et~al.}(2011{\natexlab{b}})\citenamefont{Grinstein, Kagan, Zupan, and
  Trott}}]{Grinstein:2011dz}
\bibinfo{author}{\bibfnamefont{B.}~\bibnamefont{Grinstein}},
  \bibinfo{author}{\bibfnamefont{A.~L.} \bibnamefont{Kagan}},
  \bibinfo{author}{\bibfnamefont{J.}~\bibnamefont{Zupan}}, \bibnamefont{and}
  \bibinfo{author}{\bibfnamefont{M.}~\bibnamefont{Trott}},
  \bibinfo{journal}{JHEP} \textbf{\bibinfo{volume}{1110}}, \bibinfo{pages}{072}
  (\bibinfo{year}{2011}{\natexlab{b}}), \eprint{1108.4027}.

\bibitem[{\citenamefont{Ligeti et~al.}(2011)\citenamefont{Ligeti,
  Marques~Tavares, and Schmaltz}}]{Ligeti:2011vt}
\bibinfo{author}{\bibfnamefont{Z.}~\bibnamefont{Ligeti}},
  \bibinfo{author}{\bibfnamefont{G.}~\bibnamefont{Marques~Tavares}},
  \bibnamefont{and} \bibinfo{author}{\bibfnamefont{M.}~\bibnamefont{Schmaltz}},
  \bibinfo{journal}{JHEP} \textbf{\bibinfo{volume}{1106}}, \bibinfo{pages}{109}
  (\bibinfo{year}{2011}), \eprint{1103.2757}.

\bibitem[{\citenamefont{Marques~Tavares and Schmaltz}(2011)}]{Tavares:2011zg}
\bibinfo{author}{\bibfnamefont{G.}~\bibnamefont{Marques~Tavares}}
  \bibnamefont{and} \bibinfo{author}{\bibfnamefont{M.}~\bibnamefont{Schmaltz}},
  \bibinfo{journal}{Phys.Rev.} \textbf{\bibinfo{volume}{D84}},
  \bibinfo{pages}{054008} (\bibinfo{year}{2011}), \eprint{1107.0978}.

\bibitem[{\citenamefont{Bhattacherjee et~al.}(2011)\citenamefont{Bhattacherjee,
  Biswal, and Ghosh}}]{Bhattacherjee:2011nr}
\bibinfo{author}{\bibfnamefont{B.}~\bibnamefont{Bhattacherjee}},
  \bibinfo{author}{\bibfnamefont{S.~S.} \bibnamefont{Biswal}},
  \bibnamefont{and} \bibinfo{author}{\bibfnamefont{D.}~\bibnamefont{Ghosh}},
  \bibinfo{journal}{Phys.Rev.} \textbf{\bibinfo{volume}{D83}},
  \bibinfo{pages}{091501} (\bibinfo{year}{2011}), \eprint{1102.0545}.

\bibitem[{\citenamefont{Arkani-Hamed et~al.}(1998)\citenamefont{Arkani-Hamed,
  Dimopoulos, and Dvali}}]{ArkaniHamed:1998rs}
\bibinfo{author}{\bibfnamefont{N.}~\bibnamefont{Arkani-Hamed}},
  \bibinfo{author}{\bibfnamefont{S.}~\bibnamefont{Dimopoulos}},
  \bibnamefont{and} \bibinfo{author}{\bibfnamefont{G.}~\bibnamefont{Dvali}},
  \bibinfo{journal}{Phys.Lett.} \textbf{\bibinfo{volume}{B429}},
  \bibinfo{pages}{263} (\bibinfo{year}{1998}), \eprint{hep-ph/9803315}.

\bibitem[{\citenamefont{Antoniadis et~al.}(1998)\citenamefont{Antoniadis,
  Arkani-Hamed, Dimopoulos, and Dvali}}]{Antoniadis:1998ig}
\bibinfo{author}{\bibfnamefont{I.}~\bibnamefont{Antoniadis}},
  \bibinfo{author}{\bibfnamefont{N.}~\bibnamefont{Arkani-Hamed}},
  \bibinfo{author}{\bibfnamefont{S.}~\bibnamefont{Dimopoulos}},
  \bibnamefont{and} \bibinfo{author}{\bibfnamefont{G.}~\bibnamefont{Dvali}},
  \bibinfo{journal}{Phys.Lett.} \textbf{\bibinfo{volume}{B436}},
  \bibinfo{pages}{257} (\bibinfo{year}{1998}), \eprint{hep-ph/9804398}.

\bibitem[{\citenamefont{Randall and
  Sundrum}(1999{\natexlab{a}})}]{Randall:1999ee}
\bibinfo{author}{\bibfnamefont{L.}~\bibnamefont{Randall}} \bibnamefont{and}
  \bibinfo{author}{\bibfnamefont{R.}~\bibnamefont{Sundrum}},
  \bibinfo{journal}{Phys.Rev.Lett.} \textbf{\bibinfo{volume}{83}},
  \bibinfo{pages}{3370} (\bibinfo{year}{1999}{\natexlab{a}}),
  \eprint{hep-ph/9905221}.

\bibitem[{\citenamefont{Randall and
  Sundrum}(1999{\natexlab{b}})}]{Randall:1999vf}
\bibinfo{author}{\bibfnamefont{L.}~\bibnamefont{Randall}} \bibnamefont{and}
  \bibinfo{author}{\bibfnamefont{R.}~\bibnamefont{Sundrum}},
  \bibinfo{journal}{Phys.Rev.Lett.} \textbf{\bibinfo{volume}{83}},
  \bibinfo{pages}{4690} (\bibinfo{year}{1999}{\natexlab{b}}),
  \eprint{hep-th/9906064}.

\bibitem[{\citenamefont{de~Rham and Gabadadze}(2010)}]{deRham:2010gu}
\bibinfo{author}{\bibfnamefont{C.}~\bibnamefont{de~Rham}} \bibnamefont{and}
  \bibinfo{author}{\bibfnamefont{G.}~\bibnamefont{Gabadadze}},
  \bibinfo{journal}{Phys.Lett.} \textbf{\bibinfo{volume}{B693}},
  \bibinfo{pages}{334} (\bibinfo{year}{2010}), \eprint{1006.4367}.

\bibitem[{\citenamefont{de~Rham et~al.}(2011)\citenamefont{de~Rham, Gabadadze,
  and Tolley}}]{deRham:2010kj}
\bibinfo{author}{\bibfnamefont{C.}~\bibnamefont{de~Rham}},
  \bibinfo{author}{\bibfnamefont{G.}~\bibnamefont{Gabadadze}},
  \bibnamefont{and} \bibinfo{author}{\bibfnamefont{A.~J.}
  \bibnamefont{Tolley}}, \bibinfo{journal}{Phys.Rev.Lett.}
  \textbf{\bibinfo{volume}{106}}, \bibinfo{pages}{231101}
  (\bibinfo{year}{2011}), \eprint{1011.1232}.

\bibitem[{\citenamefont{Morningstar and Peardon}(1997)}]{Morningstar:1997ff}
\bibinfo{author}{\bibfnamefont{C.~J.} \bibnamefont{Morningstar}}
  \bibnamefont{and} \bibinfo{author}{\bibfnamefont{M.~J.}
  \bibnamefont{Peardon}}, \bibinfo{journal}{Phys.Rev.}
  \textbf{\bibinfo{volume}{D56}}, \bibinfo{pages}{4043} (\bibinfo{year}{1997}),
  \eprint{hep-lat/9704011}.

\bibitem[{\citenamefont{Morningstar and Peardon}(1999)}]{Morningstar:1999rf}
\bibinfo{author}{\bibfnamefont{C.~J.} \bibnamefont{Morningstar}}
  \bibnamefont{and} \bibinfo{author}{\bibfnamefont{M.~J.}
  \bibnamefont{Peardon}}, \bibinfo{journal}{Phys.Rev.}
  \textbf{\bibinfo{volume}{D60}}, \bibinfo{pages}{034509}
  (\bibinfo{year}{1999}), \eprint{hep-lat/9901004}.

\bibitem[{\citenamefont{Chen et~al.}(2006)\citenamefont{Chen, Alexandru, Dong,
  Draper, Horvath et~al.}}]{Chen:2005mg}
\bibinfo{author}{\bibfnamefont{Y.}~\bibnamefont{Chen}},
  \bibinfo{author}{\bibfnamefont{A.}~\bibnamefont{Alexandru}},
  \bibinfo{author}{\bibfnamefont{S.}~\bibnamefont{Dong}},
  \bibinfo{author}{\bibfnamefont{T.}~\bibnamefont{Draper}},
  \bibinfo{author}{\bibfnamefont{I.}~\bibnamefont{Horvath}},
  \bibnamefont{et~al.}, \bibinfo{journal}{Phys.Rev.}
  \textbf{\bibinfo{volume}{D73}}, \bibinfo{pages}{014516}
  (\bibinfo{year}{2006}), \eprint{hep-lat/0510074}.

\bibitem[{\citenamefont{Fok et~al.}(2012)\citenamefont{Fok, Guimaraes, Lewis,
  and Sanz}}]{Fok:2012zk}
\bibinfo{author}{\bibfnamefont{R.}~\bibnamefont{Fok}},
  \bibinfo{author}{\bibfnamefont{C.}~\bibnamefont{Guimaraes}},
  \bibinfo{author}{\bibfnamefont{R.}~\bibnamefont{Lewis}}, \bibnamefont{and}
  \bibinfo{author}{\bibfnamefont{V.}~\bibnamefont{Sanz}}
  (\bibinfo{year}{2012}), \eprint{1203.2917}.

\bibitem[{\citenamefont{Fierz and Pauli}(1939)}]{Fierz:1939ix}
\bibinfo{author}{\bibfnamefont{M.}~\bibnamefont{Fierz}} \bibnamefont{and}
  \bibinfo{author}{\bibfnamefont{W.}~\bibnamefont{Pauli}},
  \bibinfo{journal}{Proc.Roy.Soc.Lond.} \textbf{\bibinfo{volume}{A173}},
  \bibinfo{pages}{211} (\bibinfo{year}{1939}).

\bibitem[{\citenamefont{Donoghue}(1994)}]{Donoghue:1994dn}
\bibinfo{author}{\bibfnamefont{J.~F.} \bibnamefont{Donoghue}},
  \bibinfo{journal}{Phys.Rev.} \textbf{\bibinfo{volume}{D50}},
  \bibinfo{pages}{3874} (\bibinfo{year}{1994}), \eprint{gr-qc/9405057}.

\bibitem[{\citenamefont{Giudice et~al.}(1999)\citenamefont{Giudice, Rattazzi,
  and Wells}}]{Giudice:1998ck}
\bibinfo{author}{\bibfnamefont{G.~F.} \bibnamefont{Giudice}},
  \bibinfo{author}{\bibfnamefont{R.}~\bibnamefont{Rattazzi}}, \bibnamefont{and}
  \bibinfo{author}{\bibfnamefont{J.~D.} \bibnamefont{Wells}},
  \bibinfo{journal}{Nucl.Phys.} \textbf{\bibinfo{volume}{B544}},
  \bibinfo{pages}{3} (\bibinfo{year}{1999}), \eprint{hep-ph/9811291}.

\bibitem[{\citenamefont{Arkani-Hamed et~al.}(2003)\citenamefont{Arkani-Hamed,
  Georgi, and Schwartz}}]{ArkaniHamed:2002sp}
\bibinfo{author}{\bibfnamefont{N.}~\bibnamefont{Arkani-Hamed}},
  \bibinfo{author}{\bibfnamefont{H.}~\bibnamefont{Georgi}}, \bibnamefont{and}
  \bibinfo{author}{\bibfnamefont{M.~D.} \bibnamefont{Schwartz}},
  \bibinfo{journal}{Annals Phys.} \textbf{\bibinfo{volume}{305}},
  \bibinfo{pages}{96} (\bibinfo{year}{2003}), \eprint{hep-th/0210184}.

\bibitem[{\citenamefont{Berger et~al.}(2012)\citenamefont{Berger, Cao, Chen,
  Yu, and Zhang}}]{Berger:2012nw}
\bibinfo{author}{\bibfnamefont{E.~L.} \bibnamefont{Berger}},
  \bibinfo{author}{\bibfnamefont{Q.-H.} \bibnamefont{Cao}},
  \bibinfo{author}{\bibfnamefont{C.-R.} \bibnamefont{Chen}},
  \bibinfo{author}{\bibfnamefont{J.-H.} \bibnamefont{Yu}}, \bibnamefont{and}
  \bibinfo{author}{\bibfnamefont{H.}~\bibnamefont{Zhang}}
  (\bibinfo{year}{2012}), \eprint{1201.1790}.

\bibitem[{\citenamefont{Grinstein
  et~al.}(2011{\natexlab{c}})\citenamefont{Grinstein, Murphy, and
  Trott}}]{Grinstein:2011gq}
\bibinfo{author}{\bibfnamefont{B.}~\bibnamefont{Grinstein}},
  \bibinfo{author}{\bibfnamefont{C.~W.} \bibnamefont{Murphy}},
  \bibnamefont{and} \bibinfo{author}{\bibfnamefont{M.}~\bibnamefont{Trott}},
  \bibinfo{journal}{JHEP} \textbf{\bibinfo{volume}{1111}}, \bibinfo{pages}{139}
  (\bibinfo{year}{2011}{\natexlab{c}}), \eprint{1110.5361}.

\bibitem[{\citenamefont{Gresham et~al.}(2012)\citenamefont{Gresham, Kim, Tulin,
  and Zurek}}]{Gresham:2012wc}
\bibinfo{author}{\bibfnamefont{M.~I.} \bibnamefont{Gresham}},
  \bibinfo{author}{\bibfnamefont{I.-W.} \bibnamefont{Kim}},
  \bibinfo{author}{\bibfnamefont{S.}~\bibnamefont{Tulin}}, \bibnamefont{and}
  \bibinfo{author}{\bibfnamefont{K.~M.} \bibnamefont{Zurek}}
  (\bibinfo{year}{2012}), \eprint{1203.1320}.

\bibitem[{\citenamefont{{Particle Data Group}}(2010)}]{Nakamura:2010zzi}
\bibinfo{author}{\bibnamefont{{Particle Data Group}}},
  \bibinfo{journal}{J.Phys.G} \textbf{\bibinfo{volume}{G37}},
  \bibinfo{pages}{075021} (\bibinfo{year}{2010}).

\bibitem[{\citenamefont{{ATLAS Collaboration}}(2012{\natexlab{a}})}]{:2012bb}
\bibinfo{author}{\bibnamefont{{ATLAS Collaboration}}}
  (\bibinfo{year}{2012}{\natexlab{a}}), \eprint{1202.5520}.

\bibitem[{\citenamefont{{CMS
  Collaboration}}(2011{\natexlab{a}})}]{Chatrchyan:2011dk}
\bibinfo{author}{\bibnamefont{{CMS Collaboration}}}, \bibinfo{journal}{JHEP}
  \textbf{\bibinfo{volume}{1108}}, \bibinfo{pages}{005}
  (\bibinfo{year}{2011}{\natexlab{a}}), \eprint{1106.2142}.

\bibitem[{\citenamefont{{UTfit Collaboration}}(2008)}]{Bona:2007vi}
\bibinfo{author}{\bibnamefont{{UTfit Collaboration}}}, \bibinfo{journal}{JHEP}
  \textbf{\bibinfo{volume}{0803}}, \bibinfo{pages}{049} (\bibinfo{year}{2008}),
  \eprint{0707.0636}.

\bibitem[{\citenamefont{{CDF
  Collaboration}}(2009{\natexlab{a}})}]{Aaltonen:2008dn}
\bibinfo{author}{\bibnamefont{{CDF Collaboration}}},
  \bibinfo{journal}{Phys.Rev.} \textbf{\bibinfo{volume}{D79}},
  \bibinfo{pages}{112002} (\bibinfo{year}{2009}{\natexlab{a}}),
  \eprint{0812.4036}.

\bibitem[{\citenamefont{{D0 Collaboration}}(2012)}]{Abazov:2012vd}
\bibinfo{author}{\bibnamefont{{D0 Collaboration}}} (\bibinfo{year}{2012}),
  \eprint{1201.4156}.

\bibitem[{\citenamefont{Manohar and Trott}(2012)}]{Manohar:2012rs}
\bibinfo{author}{\bibfnamefont{A.~V.} \bibnamefont{Manohar}} \bibnamefont{and}
  \bibinfo{author}{\bibfnamefont{M.}~\bibnamefont{Trott}}
  (\bibinfo{year}{2012}), \eprint{1201.3926}.

\bibitem[{\citenamefont{{CDF
  Collaboration}}(2011{\natexlab{c}})}]{Aaltonen:2010hza}
\bibinfo{author}{\bibnamefont{{CDF Collaboration}}},
  \bibinfo{journal}{Phys.Rev.} \textbf{\bibinfo{volume}{D83}},
  \bibinfo{pages}{071102} (\bibinfo{year}{2011}{\natexlab{c}}),
  \eprint{1007.4423}.

\bibitem[{\citenamefont{Ahrens et~al.}(2011{\natexlab{a}})\citenamefont{Ahrens,
  Ferroglia, Neubert, Pecjak, and Yang}}]{Ahrens:2011mw}
\bibinfo{author}{\bibfnamefont{V.}~\bibnamefont{Ahrens}},
  \bibinfo{author}{\bibfnamefont{A.}~\bibnamefont{Ferroglia}},
  \bibinfo{author}{\bibfnamefont{M.}~\bibnamefont{Neubert}},
  \bibinfo{author}{\bibfnamefont{B.}~\bibnamefont{Pecjak}}, \bibnamefont{and}
  \bibinfo{author}{\bibfnamefont{L.}~\bibnamefont{Yang}},
  \bibinfo{journal}{JHEP} \textbf{\bibinfo{volume}{1109}}, \bibinfo{pages}{070}
  (\bibinfo{year}{2011}{\natexlab{a}}), \eprint{1103.0550}.

\bibitem[{\citenamefont{Hollik and Pagani}(2011)}]{Hollik:2011ps}
\bibinfo{author}{\bibfnamefont{W.}~\bibnamefont{Hollik}} \bibnamefont{and}
  \bibinfo{author}{\bibfnamefont{D.}~\bibnamefont{Pagani}},
  \bibinfo{journal}{Phys.Rev.} \textbf{\bibinfo{volume}{D84}},
  \bibinfo{pages}{093003} (\bibinfo{year}{2011}), \eprint{1107.2606}.

\bibitem[{\citenamefont{Ahrens et~al.}(2011{\natexlab{b}})\citenamefont{Ahrens,
  Ferroglia, Neubert, Pecjak, and Yang}}]{Ahrens:2011uf}
\bibinfo{author}{\bibfnamefont{V.}~\bibnamefont{Ahrens}},
  \bibinfo{author}{\bibfnamefont{A.}~\bibnamefont{Ferroglia}},
  \bibinfo{author}{\bibfnamefont{M.}~\bibnamefont{Neubert}},
  \bibinfo{author}{\bibfnamefont{B.~D.} \bibnamefont{Pecjak}},
  \bibnamefont{and} \bibinfo{author}{\bibfnamefont{L.~L.} \bibnamefont{Yang}},
  \bibinfo{journal}{Phys.Rev.} \textbf{\bibinfo{volume}{D84}},
  \bibinfo{pages}{074004} (\bibinfo{year}{2011}{\natexlab{b}}),
  \eprint{1106.6051}.

\bibitem[{\citenamefont{Martin et~al.}(2009)\citenamefont{Martin, Stirling,
  Thorne, and Watt}}]{Martin:2009iq}
\bibinfo{author}{\bibfnamefont{A.}~\bibnamefont{Martin}},
  \bibinfo{author}{\bibfnamefont{W.}~\bibnamefont{Stirling}},
  \bibinfo{author}{\bibfnamefont{R.}~\bibnamefont{Thorne}}, \bibnamefont{and}
  \bibinfo{author}{\bibfnamefont{G.}~\bibnamefont{Watt}},
  \bibinfo{journal}{Eur.Phys.J.} \textbf{\bibinfo{volume}{C63}},
  \bibinfo{pages}{189} (\bibinfo{year}{2009}), \eprint{0901.0002}.

\bibitem[{\citenamefont{{D0
  Collaboration}}(2011{\natexlab{b}})}]{Abazov:2011mi}
\bibinfo{author}{\bibnamefont{{D0 Collaboration}}},
  \bibinfo{journal}{Phys.Rev.} \textbf{\bibinfo{volume}{D84}},
  \bibinfo{pages}{012008} (\bibinfo{year}{2011}{\natexlab{b}}),
  \eprint{1101.0124}.

\bibitem[{\citenamefont{Cacciari et~al.}(2011)\citenamefont{Cacciari, Czakon,
  Mangano, Mitov, and Nason}}]{Cacciari:2011hy}
\bibinfo{author}{\bibfnamefont{M.}~\bibnamefont{Cacciari}},
  \bibinfo{author}{\bibfnamefont{M.}~\bibnamefont{Czakon}},
  \bibinfo{author}{\bibfnamefont{M.~L.} \bibnamefont{Mangano}},
  \bibinfo{author}{\bibfnamefont{A.}~\bibnamefont{Mitov}}, \bibnamefont{and}
  \bibinfo{author}{\bibfnamefont{P.}~\bibnamefont{Nason}}
  (\bibinfo{year}{2011}), \eprint{1111.5869}.

\bibitem[{\citenamefont{{CDF
  Collaboration}}(2009{\natexlab{b}})}]{Aaltonen:2009iz}
\bibinfo{author}{\bibnamefont{{CDF Collaboration}}},
  \bibinfo{journal}{Phys.Rev.Lett.} \textbf{\bibinfo{volume}{102}},
  \bibinfo{pages}{222003} (\bibinfo{year}{2009}{\natexlab{b}}),
  \eprint{0903.2850}.

\bibitem[{\citenamefont{{CMS
  Collaboration}}(2012{\natexlab{a}})}]{Chatrchyan:2011hk}
\bibinfo{author}{\bibnamefont{{CMS Collaboration}}}, \bibinfo{journal}{Phys.
  Lett. B} \textbf{\bibinfo{volume}{709}}, \bibinfo{pages}{28}
  (\bibinfo{year}{2012}{\natexlab{a}}), \eprint{1112.5100}.

\bibitem[{\citenamefont{Kuhn and Rodrigo}(2012)}]{Kuhn:2011ri}
\bibinfo{author}{\bibfnamefont{J.~H.} \bibnamefont{Kuhn}} \bibnamefont{and}
  \bibinfo{author}{\bibfnamefont{G.}~\bibnamefont{Rodrigo}},
  \bibinfo{journal}{JHEP} \textbf{\bibinfo{volume}{1201}}, \bibinfo{pages}{063}
  (\bibinfo{year}{2012}), \eprint{1109.6830}.

\bibitem[{\citenamefont{{ATLAS
  Collaboration}}(2012{\natexlab{b}})}]{ATLAS:2012an}
\bibinfo{author}{\bibnamefont{{ATLAS Collaboration}}},
  \bibinfo{journal}{Eur.Phys.J.} \textbf{\bibinfo{volume}{C72}},
  \bibinfo{pages}{2039} (\bibinfo{year}{2012}{\natexlab{b}}),
  \eprint{1203.4211}.

\bibitem[{\citenamefont{{ATLAS Collaboration}}(2012{\natexlab{c}})}]{:2012kg}
\bibinfo{author}{\bibnamefont{{ATLAS Collaboration}}}
  (\bibinfo{year}{2012}{\natexlab{c}}), \eprint{1202.4892}.

\bibitem[{\citenamefont{{CMS
  Collaboration}}(2011{\natexlab{b}})}]{Chatrchyan:2011yy}
\bibinfo{author}{\bibnamefont{{CMS Collaboration}}},
  \bibinfo{journal}{Phys.Rev.} \textbf{\bibinfo{volume}{D84}},
  \bibinfo{pages}{092004} (\bibinfo{year}{2011}{\natexlab{b}}),
  \eprint{1108.3773}.

\bibitem[{\citenamefont{Aguilar-Saavedra and
  Perez-Victoria}(2011{\natexlab{a}})}]{AguilarSaavedra:2011hz}
\bibinfo{author}{\bibfnamefont{J.}~\bibnamefont{Aguilar-Saavedra}}
  \bibnamefont{and}
  \bibinfo{author}{\bibfnamefont{M.}~\bibnamefont{Perez-Victoria}},
  \bibinfo{journal}{Phys.Rev.} \textbf{\bibinfo{volume}{D84}},
  \bibinfo{pages}{115013} (\bibinfo{year}{2011}{\natexlab{a}}),
  \eprint{1105.4606}.

\bibitem[{\citenamefont{Aguilar-Saavedra and
  Perez-Victoria}(2011{\natexlab{b}})}]{AguilarSaavedra:2011ug}
\bibinfo{author}{\bibfnamefont{J.}~\bibnamefont{Aguilar-Saavedra}}
  \bibnamefont{and}
  \bibinfo{author}{\bibfnamefont{M.}~\bibnamefont{Perez-Victoria}},
  \bibinfo{journal}{JHEP} \textbf{\bibinfo{volume}{1109}}, \bibinfo{pages}{097}
  (\bibinfo{year}{2011}{\natexlab{b}}), \eprint{1107.0841}.

\bibitem[{\citenamefont{{CMS
  Collaboration}}(2012{\natexlab{b}})}]{CMS-PAS-TOP-11-030}
\bibinfo{author}{\bibnamefont{{CMS Collaboration}}}
  (\bibinfo{year}{2012}{\natexlab{b}}), \eprint{CMS-PAS-TOP-11-030}.

\bibitem[{\citenamefont{{CMS
  Collaboration}}(2012{\natexlab{c}})}]{CMS-PAS-TOP-11-013}
\bibinfo{author}{\bibnamefont{{CMS Collaboration}}}
  (\bibinfo{year}{2012}{\natexlab{c}}), \eprint{CMS-PAS-TOP-11-013}.

\bibitem[{\citenamefont{{CDF Collaboration}}(2012)}]{CDF-Note-10807}
\bibinfo{author}{\bibnamefont{{CDF Collaboration}}} (\bibinfo{year}{2012}),
  \eprint{CDF Note 10807}.

\bibitem[{\citenamefont{Christensen and Duhr}(2009)}]{Christensen:2008py}
\bibinfo{author}{\bibfnamefont{N.~D.} \bibnamefont{Christensen}}
  \bibnamefont{and} \bibinfo{author}{\bibfnamefont{C.}~\bibnamefont{Duhr}},
  \bibinfo{journal}{Comput.Phys.Commun.} \textbf{\bibinfo{volume}{180}},
  \bibinfo{pages}{1614} (\bibinfo{year}{2009}), \eprint{0806.4194}.

\bibitem[{\citenamefont{Alwall et~al.}(2011)\citenamefont{Alwall, Herquet,
  Maltoni, Mattelaer, and Stelzer}}]{Alwall:2011uj}
\bibinfo{author}{\bibfnamefont{J.}~\bibnamefont{Alwall}},
  \bibinfo{author}{\bibfnamefont{M.}~\bibnamefont{Herquet}},
  \bibinfo{author}{\bibfnamefont{F.}~\bibnamefont{Maltoni}},
  \bibinfo{author}{\bibfnamefont{O.}~\bibnamefont{Mattelaer}},
  \bibnamefont{and} \bibinfo{author}{\bibfnamefont{T.}~\bibnamefont{Stelzer}},
  \bibinfo{journal}{JHEP} \textbf{\bibinfo{volume}{1106}}, \bibinfo{pages}{128}
  (\bibinfo{year}{2011}), \eprint{1106.0522}.

\bibitem[{\citenamefont{Drobnak et~al.}(2012)\citenamefont{Drobnak, Kamenik,
  and Zupan}}]{Drobnak:2012cz}
\bibinfo{author}{\bibfnamefont{J.}~\bibnamefont{Drobnak}},
  \bibinfo{author}{\bibfnamefont{J.~F.} \bibnamefont{Kamenik}},
  \bibnamefont{and} \bibinfo{author}{\bibfnamefont{J.}~\bibnamefont{Zupan}}
  (\bibinfo{year}{2012}), \eprint{1205.4721}.

\end{thebibliography}

\end{document}